\documentclass[
  journal=,
  manuscript=,  
]{cup-journal}

\usepackage[nopatch]{microtype}
\usepackage{booktabs}
\usepackage{caption}
\usepackage{tabularray}
\usepackage{supertabular}
\usepackage{graphicx}
\usepackage{setspace}

\title{\Large\fontfamily{ptm}\selectfont Does Financial Literacy Impact Investment Participation and Retirement Planning in Japan?\footnote{Acknowledgements: This research utilized individual-level data from the “Financial Literacy Survey,” which was graciously provided by the Central Council for Financial Services Information (Secretariat: Public Relations Department, Bank of Japan).
This work was partially supported by JSPS KAKENHI 20K11708.}}

\author{Yi Jiang}
\affiliation{Graduate School of Data Science, Shiga University, Japan}
\email[Yi Jiang]{s7022101@st.shiga-u.ac.jp}

\author{Shohei Shimizu}
\affiliation{Graduate School of Data Science, Shiga University, Japan}
\alsoaffiliation{Center for Advanced Intelligence Project, RIKEN, Japan}

\usepackage{etoolbox}
\makeatletter
\patchcmd{\@afterheading}%
    {\clubpenalty \@M}{\clubpenalties 3 \@M \@M 0}{}{}
\patchcmd{\@afterheading}%
    {\@nobreaktrue}{\@nobreakfalse}{}{}
\makeatother

\begin{document}
\doublespacing


\noindent\textbf{Abstract}

\noindent{By employing causal discovery method, the Fast Causal Inference (FCI) model to analyze data from the 2022 “Financial Literacy Survey,” we explore the causal relationships between financial literacy and financial activities, specifically investment participation and retirement planning. Our findings indicate that increasing financial literacy may not directly boost engagement in financial investments or retirement planning in Japan, which underscores the necessity for alternative strategies to motivate financial activities among Japanese households. This research offers valuable insights for policymakers focused on improving financial well-being by advancing the use of causal discovery algorithms in understanding financial behaviors.}

\noindent\textbf{Keywords:} financial literacy, investment participation, retirement planning, causal discovery, Fast Causal Inference

\section{Introduction}
In 2019, Japan's Financial Services Agency (FSA) published a panel report titled “Asset Formation and Management in an Aging Society.” The report highlighted that an average older couple would need approximately 20 million yen for a 30-year post-retirement life, leading to a monthly deficit of around 50,000 yen not covered by public pension programs \citep{FSA2019a}. This revelation garnered widespread media attention and sparked significant public concern regarding the financial asset formation of households in Japan. Three years later, in 2022, the Council of New Form of Capitalism Realization, established in the Cabinet Secretariat of Japan, formally adopted the “Doubling of Asset-based Incomes Plan.” This initiative aims to double the number of Nippon Individual Savings Account (NISA) holders and cumulative purchase amounts within five years. This plan is designed to encourage Japanese households to diversify their savings into risky assets, including stocks and mutual funds, which is viewed as crucial for boosting overall economic growth and enhancing personal financial well-being.

The FSA has implemented various measures to promote the transition from savings to investment over the past 20 years. For example, from 2003 to 2013, preferential tax treatment for securities investment allowed that the tax rate on dividends and capital gains from stock or investment trusts was reduced from 20\% to 10\%. In 2014, the NISA was initiated as an investment tax-break scheme. In 2018, an installment-type NISA was launched, drawing inspiration from the investment strategies advocated by Burton Malkiel and Charles Ellis, renowned economists in the asset management field. Despite these initiatives, the impact appears to be less dramatic than anticipated. As illustrated in the “Flow of Funds for the Fourth Quarter of 2023” report by the Bank of Japan (BOJ), the proportion of each type of asset has remained relatively stable over the last two decades. At the end of December 2023, the share of “currency and deposits” in the total outstanding financial assets held by households was still as high as 52.6\% \citep{BOJ2023a}. 

Due to the prolonged ultra-low interest rate environment for Japanese Yen deposits, portfolios with a significant allocation to “currency and deposits” have experienced a notable disparity in total financial asset growth. According to the data published by FSA in 2019, in the 20-year period starting from 1998, the total amount of household financial assets in the United States (U.S.) and the United Kingdom (U.K.) expanded to 2.7 and 2.3 times their original values, while Japan's growth was limited to only 1.4 times during the same period \citep{FSA2019b}. This discrepancy is attributed significantly to the differences in investment returns among these countries.

In Japan, characterized by a rapidly aging population and a pay-as-you-go public pension system, the emphasis on financial asset accumulation and proactive retirement planning is becoming increasingly critical. Various academic studies and governmental surveys have been undertaken to elucidate the reasons behind the reluctance of Japanese households to invest in risky assets. With this context, the role of financial literacy, defined as individuals' capacity to process information and make judicious decisions about financial planning, asset accumulation, and pension-related considerations \citep{lusardi2014economic}, has emerged as a focal point. The Central Council for Financial Services Information, an organization of the BOJ responsible for conducting financial services information activities, has carried out nationwide large-scale questionnaire surveys, “Financial Literacy Survey,” in 2016, 2019, and 2022, aiming to assess the current state of financial literacy of individuals. Common selection questions were included in the questionnaire, and the percentage of correct answers on financial knowledge in Japan was lower compared to the U.S. and some other Organization for Economic Co-operation and Development (OECD) countries \citep{BOJ2023b}.

\citet{lusardi2011financiala} developed three questions as a benchmark for measuring financial literacy, based on economic models of saving and portfolio choice with the focus on three key economic concepts vital for financial decision-making: the understanding of interest compounding, inflation, and risk diversification. The design of the questions was guided by four fundamental principles: simplicity, relevance, brevity, and capacity to differentiate. This guideline was applied in the Health and Retirement Study conducted in the U.S. in 2004. The same set of questions was also incorporated into the “Financial Literacy Survey” executed by the BOJ.

Prior studies utilizing diverse survey datasets have explored the relationship between financial literacy and various financial activities. These investigations have elucidated that enhanced financial literacy contributes to more effective retirement planning \citep{lusardi2011financialb} and encourages increased investment in complex financial assets \citep{van2011financial}. Based on these research findings and the suboptimal financial literacy level among Japanese households, as revealed by the survey results, Japan revised its national school curriculum guidelines in 2022. This update mandates a heightened focus on personal finance education for high school students, with the objective of raising financial literacy within this young demographic.

While existing research on financial literacy has predominately identified positive correlational relationships with financial activities, employing causal discovery methods offers the potential to draw conclusions about causality. However, no prior research has employed causal discovery methods to work on this topic, either within Japan or internationally. Therefore, this study adopts a causal discovery approach to investigate the causal connections between financial literacy and financial activities, including retirement planning and investment participation in risky financial assets in Japan. By doing so, this research not only sheds light on understanding the impact of Japan's recent implementation of finance education in high schools but also provides a framework that can be applied to similar studies. Although our study primarily focuses on analyzing and interpreting data from Japan, the implications of applying the causal discovery approach also extend to other countries.

\section{Literature Review}
Over the last two decades, financial literacy has emerged as a key area of study, with a focus on the relationship with an individual’s capability to make informed financial decisions. Previous studies posit that financial literacy is essential in enabling individuals to engage in financial investment and aiding in effective retirement planning.

\citet{van2011financial} demonstrate that financial literacy significantly influences financial decision-making. Their analysis indicates that individuals with higher levels of financial literacy are more likely to participate in stock investment. This conclusion was derived from a regression analysis using the Ordinary Least Squares (OLS) method on data from the 2005 De Nederlandsche Banks Household Survey. \citet{thomas2018financial} applied OLS and probit models to a composite dataset from nine European countries: Austria, Belgium, Denmark, Germany, Italy, France, Switzerland, Sweden, and the Netherlands. Their study evaluated the influence of factors on the likelihood of stock market participation in 2010, affirming that financial literacy positively impacts this likelihood. The same conclusion is also applicable to Japan. \citet{yamori2022financial} conducted an online survey in 2019 on wealth building, securities, investment, and financial literacy in Japan. The analysis of this survey data, which included a straightforward regression model and an Instrumental Variables (IV) estimation, indicated that elevated financial literacy is generally linked to increased participation in the stock market. 

Other previous studies have focused on the association between financial literacy and retirement planning. \citet{lusardi2011financialb} analyzed data from the National Financial Capability Survey in the U.S., which is a comprehensive survey aimed at establishing benchmarks for key indicators of financial capability and associating these indicators with demographic, behavioral, attitudinal, and financial literacy characteristics. Utilizing OLS and an IV approach, the study suggests that financial literacy significantly influences retirement planning. This conclusion remains robust even when accounting for endogeneity and possible inaccuracies in measuring of financial literacy. \citet{sekita2011financial} leveraged the OLS and an IV estimation to analyze the January – February 2010 wave of the Survey of Living Preferences and Satisfaction. This research highlights the link between financial literacy and retirement planning, revealing that enhanced financial literacy increases the likelihood of holding a retirement savings plan.

Most previous studies have consistently identified correlational associations related to financial literacy through regression analysis. Some of these studies have attempted to address specific types of endogeneity, such as reverse causality or selection bias, by adopting an IV approach \citep{lusardi2011financialb} \citep{yamori2022financial}. However, IV methods primarily concentrate on estimating specific causal relationships between selected sets of variables without consideration of latent confounding variables, which may not be sufficiently comprehensive to uncover the overarching causal mechanisms.

\section{Data}
The “Financial Literacy Survey” is an extensive questionnaire survey to evaluate the current level of financial literacy among Japanese individuals. This online survey engaged 30,000 respondents aged between 18 and 79, selected to correspond closely with the demographic distribution in Japan.\footnote{The questions and options can be found on page 38 -- page 51 of \url{https://www.shiruporuto.jp/e/survey/kinyulite/pdf/22kinyulite.pdf}} For this study, we obtained individual-level data from the 2022 survey from the Central Council for Financial Services Information (Secretariat: Public Relations Department of the BOJ). This survey is the third installment following the conduction in 2016 and 2019. 

We used 13 variables from the survey data to represent the demographics, socioeconomic status, behavioral biases, financial activities, and the level of financial literacy of respondents. The list of variables is provided in \ref{apd:third}.

First, we identified five exogenous variables from the survey data for demographics and socioeconomic status.\\
(1) \textit{Male}: Dummy Variable. Assigned a value of 1 if the response to “Q42. What is your gender?” is “Male.”\\
(2) \textit{Fin\_Edu}: Dummy Variable. Set to 1 if the answer to “Q39. Was financial education offered by a school or college you attended, or a workplace where you were employed?” is “Yes, and I did participate in the financial education.”\\ 
(3) \textit{Fin\_Edu\_Home}: Dummy Variable. Assigned a value of 1 if the response to “Q40. Did your parents or guardians teach you how to manage your finances?” is “Yes.”\\
(4) \textit{Age}: Calculated as the mean of the age range indicated by the response to “Q43. What is your age?” For example, a response is “1,” denoting the age range of 18 -- 19, results in an assigned value of “18.5.” The age ranges corresponding to each answer option are as follows: “2” represents 20 -- 24, “3” for 25 -- 29, up to “13” for 75 -- 79.\\
(5) \textit{Education}: Determined by the total years required to complete the educational level indicated in “Q44. What is your educational background?” For instance, if “Primary and secondary schools only” is selected, then \textit{Education} is set as “9,” and if “University” is selected, then \textit{Education} is set as “16.” A missing value is assigned if the answer “Other” is selected.

Moreover, we included two more variables to capture additional information about socioeconomic status. We incorporated a variable \textit{Income} from the answer to “Q51. Which of these categories do your (your household) income for last year fall into?” We calculated the mean of the range for each category: a value of “0” for “Don’t have any financial assets,” “125” for “Less than 2.5 million yen,” “375” for “At least 2.5 million but less than 5 million yen,” “625” for “At least 5 million but less than 7.5 million yen,” “875” for “At least 7.5 million but less than 10 million yen,” and “1250” for “At least 10 million but less than 15 million yen.” For the answer “At least 15 million yen,” we assigned the lower limit value, “1500,” for \textit{Income}. A missing value is set if the response is “Don't know/Prefer not to say.” Similarly, we determined the variable \textit{Asset\_Amt} based on the answer to “Q52. Which of these categories do your (your household’s) financial assets (deposits, stocks, etc.) currently fall into?” The assigned values are “0” for “Don’t have any financial assets,” “125” for “Less than 2.5 million yen,” “375” for “At least 2.5 million but less than 5 million yen,” “625” for “At least 5 million but less than 7.5 million yen,” “875” for “At least 7.5 million but less than 10 million yen,” and “1500” for “At least 10 million but less than 20 million yen.” For “At least 20 million yen,” we assigned “2000” as the value of \textit{Asset\_Amt}. A missing value is set for the answer “Don't know/Prefer not to say.” Variables for income and asset amount are considered crucial because they have been demonstrated to be related to the level of financial literacy \citep{sekita2011financial} \citep{lusardi2019financial}.

In the 2011 report “National Financial Literacy Strategy,” the Australian Securities \& Investments Commission identified six primary behavioral biases that may influence individuals’ decisions, including those related to complex financial matters. These biases are disengagement, overconfidence, loss aversion, myopic decision-making, mental accounting, and herd behavior. For myopic decision making, which involves prioritizing short-term gains without considering future consequences, we adopted the answer to “Q1\_10. If I had the choice of (1) receiving 100,000 yen now or (2) receiving 110,000 yen in 1 year, I would choose (1), provided that I can definitely receive the money” as the indicator variable \textit{Myopic\_Bias}. Moreover, we identified herd behavior through \textit{Herding\_Bias}, based on the response to “Q1\_3. When there are several similar products, I tend to buy what is recommended as the best-selling product, rather than what I actually think is a good product.” The answer options for Q1\_10 and Q1\_3, which range from “1” (Agree) to “5” (Disagree), employ the Likert scale, a psychometric scale commonly used in questionnaires. This scale, named after the social scientist Rensis Likert, allows respondents to indicate their level of agreement or disagreement with a statement by choosing from a specified range of options. In addition, we utilized the answer to “Q17. How would you rate your overall knowledge about financial matters compared with other people?” as a measure of respondents’ confidence in their financial literacy level (\textit{Confidence}). The choices for Q17, ranging from “1” (Very high) to “5” (Very low), with “6” representing “Don’t know,” also follow the Likert scale format. A missing value is assigned to \textit{Confidence} if option “6” is selected. The combination of objective financial literacy and subjective confidence can be used to measure the individuals’ overconfidence \citep{biais2005judgemental}, and is reported to be positively correlated with stock market participation \citep{xia2014financial}. Unfortunately, for the remaining three biases, we could not identify suitable corresponding questions within the survey.

A variable \textit{Invest} was constructed to qualify the number of risky asset types an individual has invested in, based on the responses to “Q34. Have you ever purchased any of the following financial products?” Risky assets are defined as stocks, investment trusts, and foreign currency deposits/money market funds. Another variable, \textit{Planning}, was created to assess whether the individual is well prepared for retirement, according to the answers to a series of questions: “Q7. What expenses do you think you will have to cover in the future?,” “Q8\_1. Are you aware of the amounts that will be required for living expenses for retirement in your case?,” “Q9\_1. Do you have a financial plan for the expenses you think you will have to cover for living expenses for retirement?” and “Q10\_1. Have you set aside funds for the expenses you think you will have to cover for living expenses for retirement?” If the option “1. Living expenses for retirement” is chosen in Q7, 1 is added to \textit{Planning}. Additionally, for each “Yes” response to Q8\_1, Q9\_1, and Q10\_1, \textit{Planning} is incremented by 1. Therefore, \textit{Planning} can range from “0” to “4,” where a higher value indicates more comprehensive planning and preparation for retirement.

Finally, and of significant importance, \textit{Fin\_Literacy} represents the number of correct answers to the 25 questions designed to evaluate the individual’s financial literacy level, covering eight categories of the Financial Literacy Map. See \ref{apd:second} for all the question numbers and corresponding categories. Although the BOJ does not publicly disclose the correct answer to each question, we deduced them from our background knowledge in finance. Moreover, the “Financial Literacy Survey: 2022 Results” include each option's response ratio and the correct answer rate for every question \citep{BOJ2023b}. We compared the response ratios of the options deemed correct with the correct answer rates to ascertain the accuracy of our answers.

Apart from the three dummy variables -- \textit{Male}, \textit{Fin\_Edu}, \textit{Fin\_Edu\_Home} -- all other variables are treated as continuous. These continuous variables have been standardized before the subsequent analysis. We eliminated data entries with missing values for any of the variables, resulting in a reduction of the sample size from 30,000 to 19,333. This exclusion process could potentially introduce bias, particularly concerning the data entries for income and financial assets. The selection of the “Don't know/Prefer not to say” option to these questions may not occur randomly. Instead, it is conceivable that individuals with either particularly low or high income (or financial assets) might tend to choose “Prefer not to say.”

\section{Method}
To elucidate causal relationships, the traditional methods involve interventions or randomized controlled trials \citep{glymour2019review}. However, these approaches are often too expensive, time-intensive, or impractical to implement for specific fields of study. For instance, it is hardly permissible to engage in randomized A/B tests concerning financial education and financial literacy, particularly from the perspective of social ethics and fairness. Considering these challenges, causal discovery methods are gaining increasing interest across various disciplines. These methods facilitate the derivation of causal structures from observational data, offering a practical alternative to more traditional, intervention-based approaches.

As the application of causal discovery models requires establishing specific foundational assumptions \citep{heinze2018causal}, our initial focus will be on discussing the assumptions to select the most appropriate model. As noted in Section 3, our inability to locate corresponding questions within the survey that align with concepts of disengagement, loss aversion, and mental accounting – all of which have been suggested to influence an individual's financial activities – leads us to conclude that the assumption of “\textbf{\textit{causal sufficiency}}” is likely violated. The causal sufficiency assumption necessitates the absence of unobserved (or latent) variables in the analysis \citep{spirtes2000causation}. While the linear, non-Gaussian, and acyclic model (\textbf{LiNGAM}) \citep{shimizu2006linear}, as one of the structural equation models \citep{drton2017structure}, is capable of identifying the unique directed acyclic graph (DAG), it requires causal sufficiency, which is not satisfied in the context of this study. Among the constraint-based methods, the \textbf{PC} algorithm, named after its inventors Peter Spirtes and Clark Glymour \citep{spirtes2000causation}, necessitates the assumptions of acyclicity, causal faithfulness, and causal sufficiency. A first extension of the PC algorithm, the \textbf{FCI} \citep{spirtes2001anytime}, relaxes the assumptions by foregoing the necessity of causal sufficiency and permitting the presence of hidden variables \citep{glymour2019review}. Therefore, we leverage the FCI algorithm to explore the mechanism behind financial literacy and financial activities using the FCI method in Causal-learn package in Python \citep{zheng2023causal}.\footnote{See \url{https://causal-learn.readthedocs.io/en/latest/}}

Given the mixed nature of our dataset, which comprises three discrete dummy variables and ten continuous variables, we initially considered the usage of FCI with a kernel-based conditional independence test \citep{zhang2011kernel}, as this method is suited for handling the complexity of mixed data types. However, kernel methods’ computational complexity scales cubically with the sample size. Due to the constraints of computational resources and considering our substantial sample size of nearly 20,000, it is impractical to implement the FCI model with the kernel-based conditional independence test in this study. As an alternative approach, we divide our data into eight groups based on the combination of values of the three dummy variables: \textit{Male}, \textit{Fin\_Edu}, \textit{Fin\_Edu\_Home}. For each of these groups, we apply the FCI model with Fisher’s Z conditional independent test \citep{fisher1921014}, which is suitable for continuous variables. A significance level of 0.05 is selected for the individual partial correlation test in this study.

Furthermore, as previous research demonstrates that incorporating background knowledge into causal discovery enhances the accuracy of exploring causal relations and lends more realism to the interpretation of the results \citep{shen2020challenges}, we define \textit{Age} and \textit{Education} as exogenous variables when applying the FCI algorithm to the pre-segmented data of the eight groups.

Additionally, bootstrapping \citep{felsenstein1985confidence} \citep{tibshirani1993introduction} is used to confirm the statistical stability of the results from the causal discovery approach before we draw conclusions. The bootstrap method involves estimating the distribution of an estimator or a test statistic through the process of resampling from the original dataset. By resampling the data multiple times and regenerating the partial ancestral graphs (PAG), the outputs estimated from the FCI algorithm for each sample, we can estimate the bootstrap probability of the existence of each edge. This process enables the evaluation of the robustness of the PAG against variations in the data sample. In this study, we specify the number of bootstrap samplings as 100 and proceed to estimate the bootstrap probability. We exclude illustrations of edges with notably low probabilities, specifically those lower than or equal to 0.2, due to their unreliability. Instead, our focus was on edges with higher bootstrap probabilities.

\section{Results}
\begin{figure}[htbp]
\centering
\includegraphics[width=1\linewidth]{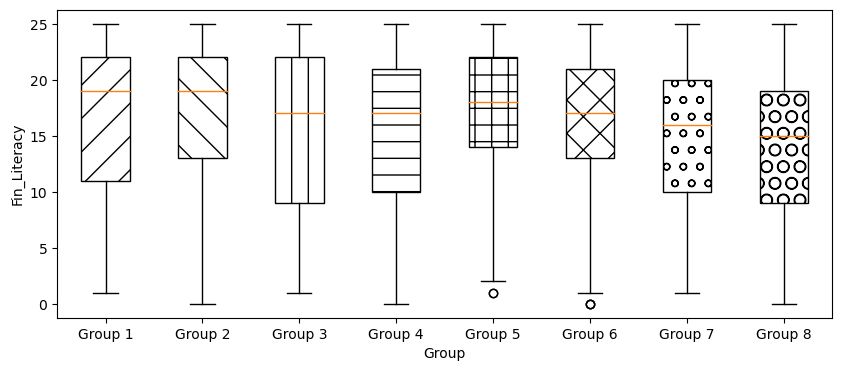}
\caption{Boxplots of \textit{“Fin\_Literacy”} for Eight Groups}
\label{boxplot}
\end{figure}

As mentioned in Section 4, we divided the original survey data into eight groups based on three dummy variables: \textit{Male}, \textit{Fin\_Edu}, and \textit{Fin\_Edu\_Home}. The sample sizes for each group, along with their corresponding conditions, are summarized in Table~\ref{data_summary}. More than 75\% of the 19,333 respondents in the research sample indicated that they did not receive financial education either at school or at home, which are segmented into Groups 4 and 8. These two groups are comparatively less financially literate, as evidenced by the number of correct answers to questions assessing financial knowledge (\textit{Fin\_Literacy)} illustrated in Figure~\ref{boxplot}.

\singlespacing
\begin{table}[htbp]
\centering
\caption{Sample Sizes of Eight Groups}
\label{data_summary}
\begin{tabular}{|c|c|c|c|c|}
\hline
\textbf{Group} & \textbf{\textit{Male}} & \textbf{\textit{Fin\_Edu}} & \textbf{\textit{Fin\_Edu\_Home}} & \textbf{Sample Size} \\\hline
Group 1 & 1 & 1 & 1 & 443 \\\hline
Group 2 & 1 & 0 & 1 & 1,482 \\\hline
Group 3 & 1 & 1 & 0 & 657 \\\hline
Group 4 & 1 & 0 & 0 & 7,961 \\\hline
Group 5 & 0 & 1 & 1 & 259 \\\hline
Group 6 & 0 & 0 & 1 & 1,659 \\\hline
Group 7 & 0 & 1 & 0 & 300 \\\hline
Group 8 & 0 & 0 & 0 & 6,572 \\\hline
\end{tabular}
\end{table}

\doublespacing
In addition, Figure~\ref{boxplot} demonstrates that female, represented by Groups 5 to 8, exhibits a lower level of financial literacy compared to male categorized by Groups 1 to 4, which aligns with the conclusion drawn from previous research \citep{lusardi2011financiala} \citep{yamori2022financial}.

\begin{figure}[hbt]
\includegraphics[width=1\linewidth]{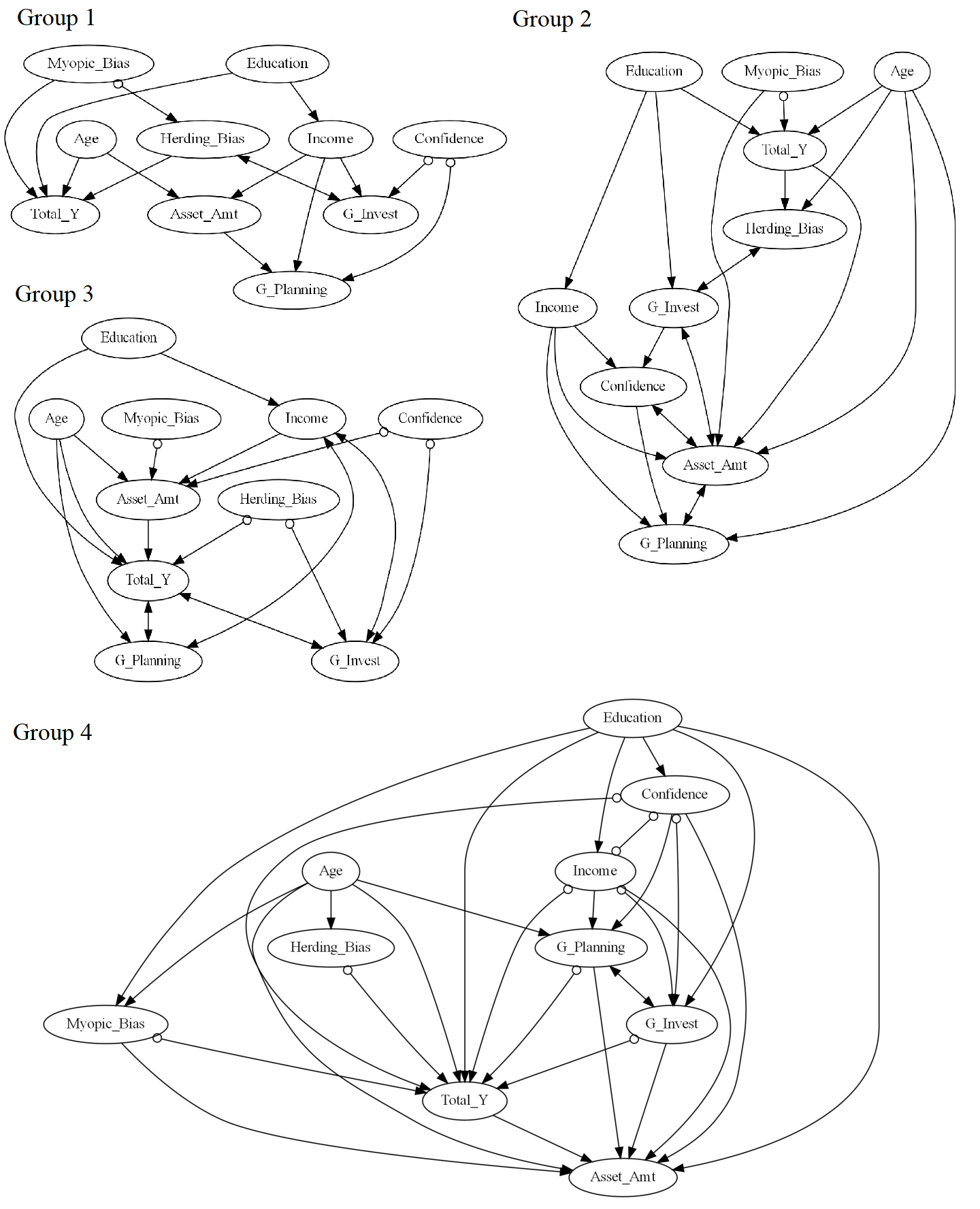}
\caption{Output PAGs of FCI for Groups 1 to 4}
\label{1_4}
\end{figure}

\begin{figure}[hbt]
\includegraphics[width=1\linewidth]{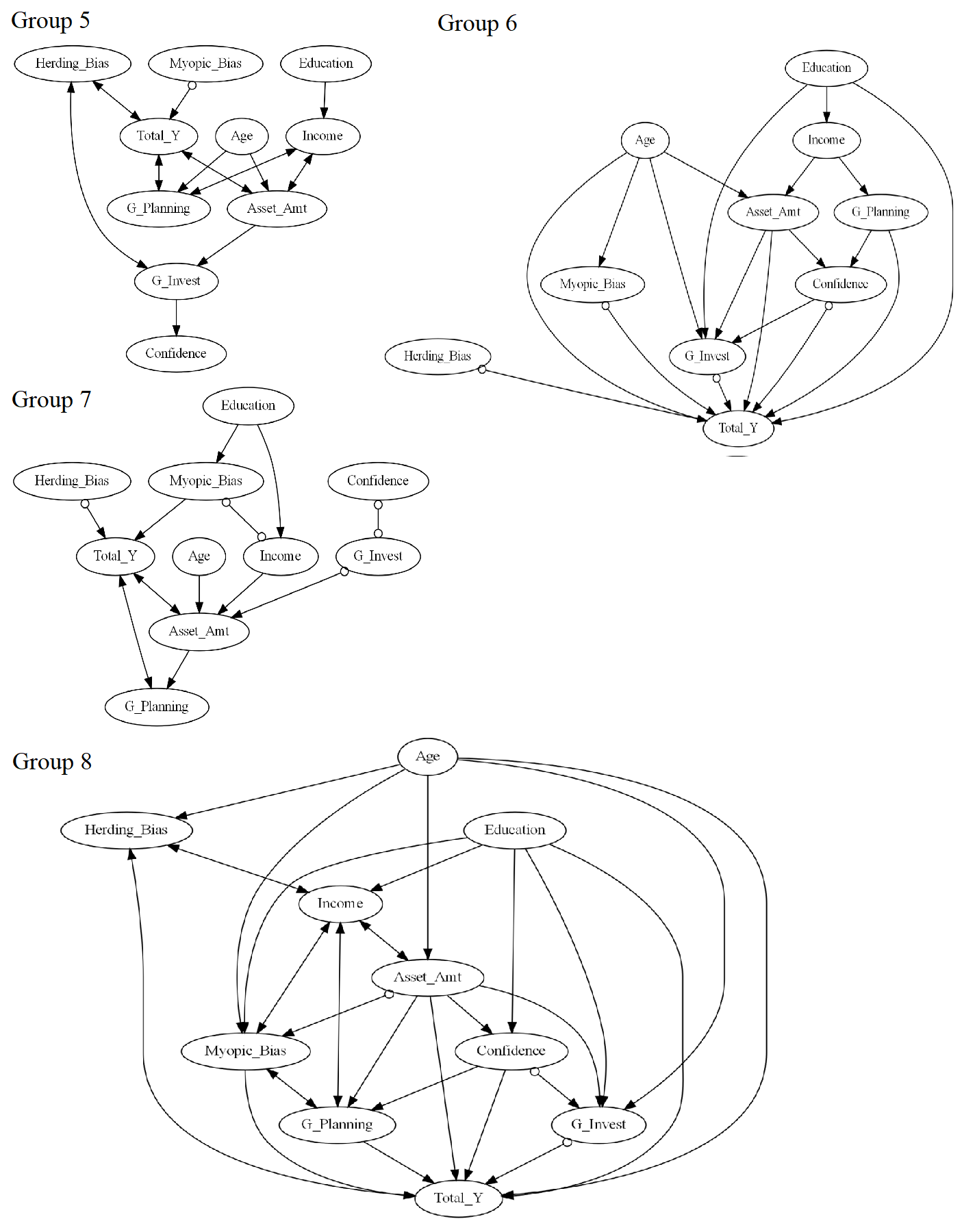}
\caption{Output PAGs of FCI for Groups 5 to 8}
\label{5_8}
\end{figure}

The PAGs derived from the FCI model, utilizing the processed survey data, are illustrated in Figure~\ref{1_4} for Groups 1 to 4 and Figure~\ref{5_8} for Groups 5 to 8. A PAG is a graphical representation used to depict the causal relationships among a set of variables, particularly when dealing with datasets that may have latent or hidden variables. In PAGs, the vertices and edges represent variables and the relationships between them. Unlike a DAG, a PAG can have four types of edges \citep{malinsky2016estimating}: $\rightarrow$, $\circ$$\rightarrow$, $\circ$–$\circ$, and $\leftrightarrow$. $i$ $\rightarrow$ $j$ is drawn when $i$ is an ancestor of $j$ and $j$ is not an ancestor of $i$. An edge $i$ $\leftrightarrow$ $j$ is drawn when $i$ is not an ancestor of $j$ and $j$ is not an ancestor of $i$, which indicates the influence of an unobserved confounder. The open dot $\circ$ is used in $i$ $\circ$$\rightarrow$ $j$ when $i$ $\rightarrow$ $j$ is present in some maximal ancestral graphs (MAG), and $i$ $\leftrightarrow$ $j$ exists in some other MAGs. 

To ascertain the reliability of the generated PAGs, we employed the bootstrap method to assess the statistical robustness of the FCI model’s outputs. The comprehensive lists of edges generated from 100 bootstrap samplings, along with their corresponding bootstrap probabilities for each group, are included in \ref{apd:fourth}. Interpretations of the findings presented in this section, as well as discussions and conclusions in subsequent sections, are based on edges with bootstrap probability larger than 0.2 in \ref{apd:fourth}.

\clearpage

\clearpage

We could not find a direct causal flow from \textit{Fin\_Literacy} to \textit{Invest} or \textit{Planning} in the PAGs of any of the eight groups, which demonstrates that the level of financial literacy is not a direct cause of participation in risky financial assets investment or retirement planning in Japan. Meanwhile, in the PAGs of Groups 1, 3, 4, 7, and 8, “$\leftrightarrow$” exists between \textit{Fin\_Literacy} and \textit{Invest}, which might indicate the existence of latent confounders. Likewise, the presence of “$\leftrightarrow$” between \textit{Fin\_Literacy} and \textit{Planning} in the PAGs of Groups 3, 4, 5, and 7 implies the possibility of unobserved variables. Moreover, “$\circ$$\rightarrow$” extends from \textit{Invest} to \textit{Fin\_Literacy} in the PAGs of Groups 4, 6, and 8. This edge suggests the potential causal relationship where participation in financial investment might enhance financial literacy levels. Similarly, “$\circ$$\rightarrow$” can be found in the PAGs of Groups 4 and 6, and “$\rightarrow$” can be found in the PAGs of Groups 6, 7, and 8 from \textit{Planning} to \textit{Fin\_Literacy}, revealing the possibility of a causal impact of retirement planning on the improvement of financial literacy.

Concerning the causal relationship about confidence in financial literacy, divergent outcomes were observed across different bootstrap samplings and groups. Specifically, for Groups 1, 2, 3, 4, 6, and 8, the PAGs of some bootstrap samples depict a possible causal flow from confidence levels to investment in risky financial assets, as represented by the “$\rightarrow$” or “$\circ$$\rightarrow$” edges from \textit{Confidence} to \textit{Invest}. In contrast, the level of confidence can be a result of participation in financial investment, illustrated by the “$\rightarrow$” from \textit{Invest} to \textit{Confidence} in some bootstrapping results of Groups 1, 2, 5, 6, and 8. Unobserved confounders can exist between \textit{Invest} and \textit{Confidence} represented by “$\leftrightarrow$” in PAGs generated from specific bootstrap samples in Groups 3 and 8.
Regarding the causal link between confidence and retirement planning, “$\rightarrow$” from \textit{Confidence} to \textit{Planning} and a reverse causal flow “$\rightarrow$” from \textit{Planning} to \textit{Confidence} can be both observed in the PAGs emerged from different bootstrap samples of Groups 1, 2, 6, and 8. For Group 4, “$\rightarrow$” and “$\leftrightarrow$” are both possible between \textit{Confidence} and \textit{Planning} in the results, indicating the possible causal flow from confidence to retirement planning and the potential latent confounders between them. “$\leftrightarrow$” between \textit{Confidence} and \textit{Planning} can also be found in Groups 2, 6, and 8 as a signal of latent variables.
 
Furthermore, beyond the scope of our initial research question related to financial literacy, the results uncovered the causal influence of age on retirement planning, which is reflected by the indirect “$\rightarrow$” from \textit{Age} to \textit{Planning} in the PAG of Groups 1, 4, 7, and 8, and direct “$\rightarrow$” in Groups 2, 3, 4, 5, and 7. Additionally, direct causal influences from age (\textit{Age}) to participation in risky asset investment (\textit{Invest}) are confirmed in the results of female Groups 5, 6, and 8. 

\section{Discussion}

Although the previous research introduced in Section 2 demonstrates that the financial literacy level is highly associated with financial activities, which could be a causal relationship, we did not confirm the direct causal impact of financial literacy on either investment in risky financial assets or retirement planning in the results of the FCI model. Therefore, enhancing the financial literacy level of individuals in Japan through mandatory financial education in high school may not inevitably improve participation in financial investment and retirement planning in Japan. Meanwhile, we found the possibility of latent confounders between financial literacy and financial activities, either participation in financial investment or retirement planning, for Groups 1, 3, 4, 5, 7, and 8. Moreover, in Groups 4, 6, 7, and 8, financial activities can have a causal impact on financial literacy levels. The existence of latent confounders and reverse causation may explain the high correlation between the variables verified in previous research.

When it comes to confidence in financial literacy, although results differed across the bootstrap samples and groups, it was a cause of investment participation in Groups 1, 2, 3, 4, 6, and 8 and of retirement planning in Groups 1, 2, 4, 6, and 8. As depicted in the “Financial Literacy Survey: 2022 Results,” a mere 12.1\% of respondents expressed confidence in their financial knowledge by rating their overall knowledge about financial matters as “Very high” or “Quite high” compared with others. This figure is significantly lower than the 76\% reported in the U.S. Subsequently, interventions, such as providing opportunities for simulated financial experiences and creating environments conducive to discussions about finances, to increase confidence in financial literacy among Japanese households, may potentially boost investment participation and retirement planning.

While the primary focus of this study remains on the impact of financial literacy, the results indicate a direct or indirect causal effect of age on financial activities. \citet{hsu2016aging} has noted that women often enhance their financial knowledge when it becomes relevant, typically around the time of their spouses’ passing. In a parallel manner, the influence of age on retirement planning and investment participation might stem from a sense of “relevancy.” As individuals age, they may perceive an escalating need for financial planning. Women may start investing in riskier assets to pursue higher returns, as their income growth does not typically match the pace experienced by men in Japan. 
	
To extend and improve this study further, variables representing disengagement, loss aversion, and mental accounting can be included in the dataset, as these factors may affect individuals’ decision-making. Given that the PAGs imply the presence of latent confounders between financial literacy and financial activities, potential variables can be identified by delving into financial and economic theories. Subsequently, redesigning the survey questions and conducting a subsequent survey are necessary for acquiring the data for further research.

Moreover, broadening the definition of financial literacy may provide deeper insights into its causal influence on financial activities. Traditionally, financial literacy has been primarily quantified in terms of financial knowledge. The practical application of this knowledge in decision-making processes was not a major focus until the OECD redefined financial literacy \citep{atkinson2012measuring}. This redefinition encompasses not only financial knowledge but also financial attitude and behavior. Therefore, financial behavior and attitude could be considered integral components of financial literacy. \citet{kadoya2020financial} suggest that financial behavior, reflecting individuals’ actions in financial matters, can be accessed by the extent of agreement with the statements below from the “Financial Literacy Survey.”\\
\indent Q1\_1. Before I buy something I carefully consider whether I can afford it.\\
\indent Q1\_2. I pay my bills on time.\\
\indent Q1\_4. I set long-term financial goals and strive to achieve them.\\
\indent Q1\_7. I keep a close personal watch on my financial affairs.\\
In a similar fashion, financial attitude, indicating one's perspective on financial issues, can be measurable through responses to the following statements.\\
\indent Q1\_5. I find it more satisfying to spend money than to save it for the long term.\\
\indent Q1\_6. I tend to live for today and let tomorrow take care of itself.

Additionally, future studies can apply the same method to the data from the 2019 iteration of the “Financial Literacy Survey.” As highlighted in Section 1, the “20-million-yen problem” garnered significant attention in 2019, intensifying public focus on financial asset formation in Japan. The onset of the COVID-19 pandemic from 2020 onwards likely influenced individuals' lifestyles, consumption habits, and financial activities. Comparing the 2019 survey results with this study could elucidate any shifts in the underlying mechanism of households’ financial activities in Japan.

\section{Conclusion}

In this research, we applied the FCI model with Fisher’s Z conditional independent test to the individual-level data from the 2022 “Financial Literacy Survey” conducted by the BOJ. This approach was employed to investigate the causal relationships between financial literacy and financial activities, specifically focusing on retirement planning and investment in risky financial assets in Japan. The original dataset was divided into eight groups based on three exogenous discrete variables to facilitate the use of Fisher’s Z conditional independent test, which is appropriate for continuous variables. Additionally, two other variables were designated as exogenous, and this information was incorporated as background knowledge about the edges. The robustness of the edges in PAGs derived from the FCI model was verified by estimating their probabilities using the bootstrap method.

While financial education at school and home may correlate with higher financial literacy, higher financial literacy does not necessarily lead to increased participation in financial investment and retirement planning in Japan. Our results did not identify direct causal links from financial literacy to financial activities. Consequently, the Japanese government may need to consider supplementary or alternative measures to achieve its objective of “a shift from savings to investment.” As confidence in financial literacy could support the adoption of desirable financial activities, initiatives can be undertaken to enhance confidence levels. The impact of age on financial activities could be attributed to perceived relevancy. Although age itself is an exogenous factor and cannot be intervened, measures could be implemented to strengthen the sense of relevancy.

We hope this study encourages further discussions on applying causal discovery algorithms to financial literacy and financial activity survey data. Equally important, we hope that the insights derived from this study will help governmental policymakers in their efforts to promote the financial well-being of households in Japan.
 
\newpage

\printbibliography

@article{lusardi2014economic,
  title={The economic importance of financial literacy: Theory and evidence},
  author={Lusardi, Annamaria and Mitchell, Olivia S},
  journal={Journal of Economic Literature},
  volume={52},
  number={1},
  pages={5--44},
  year={2014},
  publisher={American Economic Association 2014 Broadway, Suite 305, Nashville, TN 37203-2425}
}

@article{lusardi2011financiala,
  title={Financial literacy around the world: an overview},
  author={Lusardi, Annamaria and Mitchell, Olivia S},
  journal={Journal of Pension Economics \& Finance},
  volume={10},
  number={4},
  pages={497--508},
  year={2011},
  publisher={Cambridge University Press}
}

@article{lusardi2011financialb,
  title={Financial literacy and retirement planning in the United States},
  author={Lusardi, Annamaria and Mitchell, Olivia S},
  journal={Journal of Pension Economics \& Finance},
  volume={10},
  number={4},
  pages={509--525},
  year={2011},
  publisher={Cambridge University Press}
}

@article{van2011financial,
  title={Financial literacy and stock market participation},
  author={Van Rooij, Maarten and Lusardi, Annamaria and Alessie, Rob},
  journal={Journal of Financial Economics},
  volume={101},
  number={2},
  pages={449--472},
  year={2011},
  publisher={Elsevier}
}

@article{thomas2018financial,
  title={Financial literacy, human capital and stock market participation in {E}urope},
  author={Thomas, Ashok and Spataro, Luca},
  journal={Journal of Family and Economic Issues},
  volume={39},
  pages={532--550},
  year={2018},
  publisher={Springer}
}

@article{sekita2011financial,
  title={Financial literacy and retirement planning in {J}apan},
  author={Sekita, Shizuka},
  journal={Journal of Pension Economics \& Finance},
  volume={10},
  number={4},
  pages={637--656},
  year={2011},
  publisher={Cambridge University Press}
}

@article{yamori2022financial,
  title={Financial literacy and low stock market participation of {J}apanese households},
  author={Yamori, Nobuyoshi and Ueyama, Hitoe},
  journal={Finance Research Letters},
  volume={44},
  pages={102074},
  year={2022},
  publisher={Elsevier}
}

@article{biais2005judgemental,
  title={Judgemental overconfidence, self-monitoring, and trading performance in an experimental financial market},
  author={Biais, Bruno and Hilton, Denis and Mazurier, Karine and Pouget, S{\'e}bastien},
  journal={The Review of economic studies},
  volume={72},
  number={2},
  pages={287--312},
  year={2005},
  publisher={Wiley-Blackwell}
}

@article{xia2014financial,
  title={Financial literacy overconfidence and stock market participation},
  author={Xia, Tian and Wang, Zhengwei and Li, Kunpeng},
  journal={Social Indicators Research},
  volume={119},
  pages={1233--1245},
  year={2014},
  publisher={Springer}
}

@article{lusardi2019financial,
  title={Financial literacy and the need for financial education: evidence and implications},
  author={Lusardi, Annamaria},
  journal={Swiss Journal of Economics and Statistics},
  volume={155},
  number={1},
  year={2019},
  publisher={Springer}
}

@article{glymour2019review,
  title={Review of causal discovery methods based on graphical models},
  author={Glymour, Clark and Zhang, Kun and Spirtes, Peter},
  journal={Frontiers in Genetics},
  volume={10},
  pages={524},
  year={2019},
  publisher={Frontiers Media SA}
}

@article{heinze2018causal,
  title={Causal structure learning},
  author={Heinze-Deml, Christina and Maathuis, Marloes H. and Meinshausen, Nicolai},
  journal={Annual Review of Statistics and Its Application},
  volume={5},
  pages={371--391},
  year={2018},
  publisher={Annual Reviews}
}

@book{spirtes2000causation,
  title={Causation, Prediction, and Search},
  author={Spirtes, Peter and Glymour, Clark N. and Scheines, Richard},
  year={2001},
  publisher={MIT Press}
}

@article{drton2017structure,
  title={Structure learning in graphical modeling},
  author={Drton, Mathias and Maathuis, Marloes H.},
  journal={Annual Review of Statistics and Its Application},
  volume={4},
  pages={365--393},
  year={2017},
  publisher={Annual Reviews}
}

@article{shimizu2006linear,
  title={A linear non-Gaussian acyclic model for causal discovery},
  author={Shimizu, Shohei and Hoyer, Patrik O. and Hyv{\"a}rinen, Aapo and Kerminen, Antti},
  journal={Journal of Machine Learning Research},
  volume={7},
  pages={2003--2030},
  year={2006}
}

@article{fisher1921014,
  title={On the "Probable Error" of a Coefficient of Correlation Deduced from a Small Sample},
  author={Fisher, Ronald Aylmer},
  journal={Metron},
  volume={1},
  pages={1-32},
  year={1921}
}

@inproceedings{zhang2011kernel,
  title={Kernel-based Conditional Independence Test and Application in Causal Discovery},
  author={Zhang, Kun and Peters, Jonas and Janzing, Dominik and Sch{\"o}lkopf, Bernhard},
  booktitle={27th {C}onference on {U}ncertainty in {A}rtificial {I}ntelligence ({UAI} 2011)},
  pages={804--813},
  year={2011},
  organization={AUAI Press}
}

@article{shen2020challenges,
  title={Challenges and opportunities with causal discovery algorithms: Application to {A}lzheimer’s pathophysiology},
  author={Shen, Xinpeng and Ma, Sisi and Vemuri, Prashanthi and Simon, Gyorgy},
  journal={Scientific Reports},
  volume={10},
  pages={2975},
  year={2020},
  publisher={Nature Publishing Group UK London}
}

@inproceedings{malinsky2016estimating,
  title={Estimating causal effects with ancestral graph Markov models},
  author={Malinsky, Daniel and Spirtes, Peter},
  booktitle={Proceedings of the {E}ighth {I}nternational {C}onference on {P}robabilistic {G}raphical {M}odels},
  pages={299--309},
  year={2016},
  organization={PMLR}
}

@article{atkinson2012measuring,
  title={Measuring financial literacy: Results of the {OECD}/International Network on Financial Education ({INFE}) pilot study},
  author={Atkinson, Adele and Messy, Flore-Anne},
  journal={OECD Working Papers on Finance, Insurance and Private Pensions},
  volume={15},
  year={2012},
  publisher={OECD Publishing}
}

@article{hsu2016aging,
  title={Aging and strategic learning: The impact of spousal incentives on financial literacy},
  author={Hsu, Joanne W},
  journal={Journal of Human Resources},
  volume={51},
  number={4},
  pages={1036--1067},
  year={2016},
  publisher={University of Wisconsin Press}
}

@article{kadoya2020financial,
  title={Financial literacy in {J}apan: New evidence using financial knowledge, behavior, and attitude},
  author={Kadoya, Yoshihiko and Khan, Mostafa Saidur Rahim},
  journal={Sustainability},
  volume={12},
  number={9},
  pages={3683},
  year={2020},
  publisher={MDPI}
}

@article{felsenstein1985confidence,
  title={Confidence limits on phylogenies: an approach using the bootstrap},
  author={Felsenstein, Joseph},
  journal={Evolution},
  volume={39},
  number={4},
  pages={783--791},
  year={1985},
  publisher={Blackwell Publishing Inc Malden, USA}
}

@book{tibshirani1993introduction,
  title={An Introduction to the Bootstrap},
  author={Efron, Bradley and Tibshirani, Robert J},
  year={1994},
  publisher={Chapman \& Hall/CRC}
}

@article{zheng2023causal,
  title={Causal-learn: Causal discovery in python},
  author={Zheng, Yujia and Huang, Biwei and Chen, Wei and Ramsey, Joseph and Gong, Mingming and Cai, Ruichu and Shimizu, Shohei and Spirtes, Peter and Zhang, Kun},
  journal={arXiv preprint arXiv:2307.16405},
  year={2023}
}

@inproceedings{spirtes2001anytime,
  title={An anytime algorithm for causal inference},
  author={Spirtes, Peter},
  booktitle={Proceedings of the {E}ighth {I}nternational {W}orkshop on {A}rtificial {I}ntelligence and {S}tatistics},
  pages={278--285},
  volume ={R3},
  year={2001},
  organization={PMLR}
}

@article{FSA2019a,
  title={Asset Formation and Management in an Aging Society},
  author={{Financial Services Agency of Japan}},
  journal={Financial System Council, Market Working Group Report},
  year={2019}
}

@article{BOJ2023a,
  title={Basic Figures: Flow of Funds for the Fourth Quarter of 2023 (Preliminary report)},
  author={{Bank of Japan}},
  journal={Research and Statistics Department},
  year={2023}
}

@article{FSA2019b,
  title={Asset Formation in the Era of 100-year Lifespans},
  author={{Financial Services Agency of Japan}},
  journal={Financial System Council, Market Working Group Report},
  year={2019}
}

@article{BOJ2023b,
  title={Financial Literacy Survey: 2022 Results},
  author={{Bank of Japan}},
  journal={The Central Council for Financial Services Information},
  year={2022}
}
\clearpage

\singlespacing
\appendix
\setlength{\parindent}{0pt}

\section{Question Numbers for Category of Financial Literacy Map}\label{apd:second}
\vspace{\baselineskip}

\begin{supertabular}{|p{6.5cm}|p{6.2cm}|}
\hline
\textbf{Category of Financial Literacy Map} & \textbf{Question Numbers} \\ \hline
Family budget management & Q4, Q5 \\ \hline
Life planning & Q12, Q13 \\ \hline
Financial knowledge & \\ \hline
\quad Basics of financial transactions & Q14, Q15, Q16 \\ \hline
\quad Basics of finance and economy & Q18, Q19, Q20, Q21\_1, Q22, Q23 \\ \hline
\quad Insurance & Q25, Q26, Q28 \\ \hline
\quad Loans/credit & Q21\_2, Q30, Q31 \\ \hline
\quad Wealth building & Q21\_3, Q21\_4, Q33 \\ \hline
Use of outside expertise & Q36, Q37, Q38 \\ \hline
\end{supertabular}
\vspace{\baselineskip}

\doublespacing
\section{List of Variables}\label{apd:third}

\textbf{1. \textit{Male}:} Dummy Variable. Assigned a value of 1 if the
  response to “Q42. What is our gender?” is “Male.”\\
\textbf{2. \textit{Fin\_Edu}:} Dummy Variable. Set to 1 if the answer to “Q39. Was financial education offered by a school or college you attended, or a workplace where you were employed?” is “Yes, and I did participate in the financial education.” \\
\textbf{3. \textit{Fin\_Edu\_Home}:} Dummy Variable. Assigned a value of 1 if the response to “Q40. Did your parents or guardians teach you how to manage your finances?” is “Yes.” \\
\textbf{4. \textit{Age}:} Continuous Variable. Calculated as the mean of the age range indicated by the response to “Q43. What is your age?”\\
\textbf{5. \textit{Education}:} Continuous Variable. Determined by the total years required to complete the educational level indicated in “Q44. What is your educational background?”\\
\textbf{6. \textit{Income}:} Continuous Variable. The answer to “Q51. Which of these categories do your (your household) income for last year fall into?”\\
\textbf{7. \textit{Asset\_Amt}:} Continuous Variable. The answer to “Q52. Which of these categories do your (your household’s) financial assets (deposits, stocks, etc.)~ currently fall into?”\\
\textbf{8. \textit{Myopic\_Bias}:} Continuous Variable. The answer to “Q1 10. If I had the choice of (1) receiving 100,000 yen now or (2) receiving 110,000 yen in 1 year, I would choose (1), provided that I can definitely receive the money.”\\
\textbf{9. \textit{Herding\_Bias}:} Continuous Variable. The response to “Q1 3. When there are several similar products, I tend to buy what is recommended as the best-selling product, rather than what I actually think is a good product.”\\
\textbf{10. \textit{Confidence}:} Continuous Variable. The answer to “Q17. How would you rate your overall knowledge about financial matters compared with other people?”\\
\textbf{11. \textit{Invest}:} Continuous Variable. The number of risky asset types an individual has invested in, based on the responses to “Q34. Have you ever purchased any of the following financial products?”\\
\textbf{12. \textit{Planning}:} Continuous Variable. Assessed whether the individual is well prepared for retirement, according to the answers to a series of questions: Q7, Q8\_1, Q9\_1, and Q10\_1.\\
\textbf{13. \textit{Fin\_Literacy}:} Continuous Variable. The number of correct answers to the 25 questions designed to evaluate the individual’s financial literacy level.

\singlespacing
\section{Edges with Bootstrap Probabilities}\label{apd:fourth}
\twocolumn
\fontsize{10pt}{12pt}\selectfont
\tablefirsthead{%
    \hline
    \multicolumn{2}{|c|}{\textbf{Group 1}} \\ 
    \hline
    \textbf{Edge} & \textbf{Prob.} \\ 
    \hline
}
\tablehead{%
    \hline
    \multicolumn{2}{|c|}{\textbf{Group 1 (Continued)}} \\ 
    \hline
    \textbf{Edge} & \textbf{Prob.} \\
    \hline
}
\tabletail{%
    \hline
    \multicolumn{2}{|r|}{\textit{To be Continued}} \\
    \hline
}
\tablelasttail{\hline}

\begin{supertabular}{|p{5cm}|p{1cm}|}
Age $\rightarrow$ Asset\_Amt ~& 1\\ \hline
Age $\rightarrow$ Fin\_Literacy ~& 1\\ \hline
Age $\rightarrow$ Herding\_Bias ~& 0.05\\ \hline
Age $\rightarrow$ Income ~& 0.22\\ \hline
Age $\rightarrow$ Planning ~& 0.02\\ \hline
Asset\_Amt $\rightarrow$ Confidence ~& 0.11\\ \hline
Asset\_Amt $\rightarrow$ Fin\_Literacy ~& 0.12\\ \hline
Asset\_Amt $\circ$$\rightarrow$  Fin\_Literacy ~& 0.07\\ \hline
Asset\_Amt $\rightarrow$ Income ~& 0.34\\ \hline
Asset\_Amt $\circ$$\rightarrow$  Income ~& 0.01\\ \hline
Asset\_Amt $\rightarrow$ Invest ~& 0.05\\ \hline
Asset\_Amt $\leftrightarrow$ Invest ~& 0.03\\ \hline
Asset\_Amt $\rightarrow$ Planning ~& 0.58\\ \hline
Asset\_Amt $\leftrightarrow$ Planning ~& 0.16\\ \hline
Confidence $\rightarrow$ Asset\_Amt ~& 0.12\\ \hline
Confidence $\leftrightarrow$ Asset\_Amt ~& 0.08\\ \hline
Confidence $\circ$$\rightarrow$  Asset\_Amt ~& 0.06\\ \hline
Confidence $\circ$$\rightarrow$  Herding\_Bias ~& 0.01\\ \hline
Confidence $\rightarrow$ Income ~& 0.05\\ \hline
Confidence $\leftrightarrow$ Income ~& 0.02\\ \hline
Confidence $\circ$–$\circ$ Income ~& 0.01\\ \hline
Confidence $\rightarrow$ Invest ~& 0.5\\ \hline
Confidence $\circ$$\rightarrow$  Invest ~& 0.16\\ \hline
Confidence $\leftrightarrow$ Invest ~& 0.06\\ \hline
Confidence $\circ$–$\circ$ Invest ~& 0.01\\ \hline
Confidence $\rightarrow$ Planning ~& 0.37\\ \hline
Confidence $\leftrightarrow$ Planning ~& 0.16\\ \hline
Confidence $\circ$$\rightarrow$  Planning ~& 0.11\\ \hline
Confidence $\circ$–$\circ$ Planning ~& 0.01\\ \hline
Education $\rightarrow$ Asset\_Amt ~& 0.18\\ \hline
Education $\rightarrow$ Confidence ~& 0.33\\ \hline
Education $\rightarrow$ Fin\_Literacy ~& 0.69\\ \hline
Education $\rightarrow$ Income ~& 0.64\\ \hline
Education $\rightarrow$ Invest ~& 0.35\\ \hline
Education $\rightarrow$ Myopic\_Bias ~& 0.07\\ \hline
Education $\rightarrow$ Planning ~& 0.01\\ \hline
Fin\_Literacy $\leftrightarrow$ Asset\_Amt ~& 0.23\\ \hline
Fin\_Literacy $\rightarrow$ Asset\_Amt ~& 0.1\\ \hline
Fin\_Literacy $\circ$$\rightarrow$  Asset\_Amt ~& 0.01\\ \hline
Fin\_Literacy $\rightarrow$ Herding\_Bias ~& 0.21\\ \hline
Fin\_Literacy $\circ$$\rightarrow$  Herding\_Bias ~& 0.01\\ \hline
Fin\_Literacy $\leftrightarrow$ Invest ~& 0.35\\ \hline
Fin\_Literacy $\rightarrow$ Myopic\_Bias ~& 0.12\\ \hline
Fin\_Literacy $\leftrightarrow$ Planning ~& 0.05\\ \hline
Herding\_Bias $\leftrightarrow$ Confidence ~& 0.03\\ \hline
Herding\_Bias $\circ$$\rightarrow$  Confidence ~& 0.01\\ \hline
Herding\_Bias $\circ$$\rightarrow$  Fin\_Literacy ~& 0.41\\ \hline
Herding\_Bias $\rightarrow$ Fin\_Literacy ~& 0.23\\ \hline
Herding\_Bias $\leftrightarrow$ Fin\_Literacy ~& 0.14\\ \hline
Herding\_Bias $\leftrightarrow$ Invest ~& 0.31\\ \hline
Herding\_Bias $\circ$$\rightarrow$  Invest ~& 0.09\\ \hline
Herding\_Bias $\rightarrow$ Invest ~& 0.03\\ \hline
Herding\_Bias $\circ$–$\circ$ Myopic\_Bias ~& 0.22\\ \hline
Herding\_Bias $\leftrightarrow$ Myopic\_Bias ~& 0.21\\ \hline
Herding\_Bias $\circ$$\rightarrow$  Myopic\_Bias ~& 0.11\\ \hline
Herding\_Bias $\rightarrow$ Myopic\_Bias ~& 0.04\\ \hline
Herding\_Bias $\leftrightarrow$ Planning ~& 0.01\\ \hline
Income $\rightarrow$ Asset\_Amt ~& 0.59\\ \hline
Income $\circ$–$\circ$ Asset\_Amt ~& 0.03\\ \hline
Income $\circ$$\rightarrow$  Asset\_Amt ~& 0.03\\ \hline
Income $\rightarrow$ Confidence ~& 0.09\\ \hline
Income $\rightarrow$ Invest ~& 0.54\\ \hline
Income $\leftrightarrow$ Invest ~& 0.11\\ \hline
Income $\circ$$\rightarrow$  Invest ~& 0.09\\ \hline
Income $\circ$–$\circ$ Invest ~& 0.02\\ \hline
Income $\rightarrow$ Planning ~& 0.24\\ \hline
Income $\leftrightarrow$ Planning ~& 0.18\\ \hline
Income $\circ$$\rightarrow$  Planning ~& 0.04\\ \hline
Invest $\rightarrow$ Asset\_Amt ~& 0.07\\ \hline
Invest $\rightarrow$ Confidence ~& 0.27\\ \hline
Invest $\rightarrow$ Fin\_Literacy ~& 0.01\\ \hline
Invest $\rightarrow$ Herding\_Bias ~& 0.01\\ \hline
Invest $\rightarrow$ Income ~& 0.04\\ \hline
Invest $\circ$$\rightarrow$  Income ~& 0.01\\ \hline
Invest $\circ$$\rightarrow$  Planning ~& 0.01\\ \hline
Myopic\_Bias $\leftrightarrow$ Asset\_Amt ~& 0.03\\ \hline
Myopic\_Bias $\rightarrow$ Asset\_Amt ~& 0.01\\ \hline
Myopic\_Bias $\circ$$\rightarrow$  Fin\_Literacy ~& 0.47\\ \hline
Myopic\_Bias $\rightarrow$ Fin\_Literacy ~& 0.23\\ \hline
Myopic\_Bias $\leftrightarrow$ Fin\_Literacy ~& 0.15\\ \hline
Myopic\_Bias $\circ$$\rightarrow$  Herding\_Bias ~& 0.23\\ \hline
Myopic\_Bias $\rightarrow$ Herding\_Bias ~& 0.13\\ \hline
Myopic\_Bias $\leftrightarrow$ Income ~& 0.02\\ \hline
Myopic\_Bias $\circ$$\rightarrow$  Invest ~& 0.01\\ \hline
Myopic\_Bias $\leftrightarrow$ Planning ~& 0.19\\ \hline
Myopic\_Bias $\rightarrow$ Planning ~& 0.02\\ \hline
Myopic\_Bias $\circ$$\rightarrow$  Planning ~& 0.01\\ \hline
Planning $\rightarrow$ Asset\_Amt ~& 0.06\\ \hline
Planning $\circ$$\rightarrow$  Asset\_Amt ~& 0.03\\ \hline
Planning $\rightarrow$ Confidence ~& 0.22\\ \hline
Planning $\circ$$\rightarrow$  Confidence ~& 0.04\\ \hline
Planning $\rightarrow$ Fin\_Literacy ~& 0.01\\ \hline
Planning $\rightarrow$ Income ~& 0.06\\ \hline
Planning $\circ$$\rightarrow$  Income ~& 0.02\\ \hline
Planning $\leftrightarrow$ Invest ~& 0.15\\ \hline
Planning $\rightarrow$ Invest ~& 0.09\\ \hline
Planning $\circ$$\rightarrow$  Invest ~& 0.03\\ \hline

\end{supertabular}

\vspace{20pt} 

\tablefirsthead{\hline
    \multicolumn{2}{|c|}{\textbf{Group 2}} \\
    \hline
    \textbf{Edge} & \textbf{Prob.} \\ 
    \hline
}
\tablehead{\hline
    \multicolumn{2}{|c|}{\textbf{Group 2 (Continued)}} \\
    \hline
    \textbf{Edge} & \textbf{Prob.} \\
    \hline
}
\tabletail{
    \hline
    \multicolumn{2}{|r|}{\textit{To be Continued}} \\
    \hline
}
\tablelasttail{\hline}

\begin{supertabular}{|p{5cm}|p{1cm}|}
Age $\rightarrow$ Asset\_Amt ~& 1\\ \hline
Age $\rightarrow$ Fin\_Literacy ~& 1\\ \hline
Age $\rightarrow$ Herding\_Bias ~& 0.98\\ \hline
Age $\rightarrow$ Income ~& 0.06\\ \hline
Age $\rightarrow$ Myopic\_Bias ~& 0.43\\ \hline
Age $\rightarrow$ Planning ~& 0.96\\ \hline
Asset\_Amt $\rightarrow$ Confidence ~& 0.3\\ \hline
Asset\_Amt $\rightarrow$ Fin\_Literacy ~& 0.22\\ \hline
Asset\_Amt $\rightarrow$ Income ~& 0.22\\ \hline
Asset\_Amt $\leftrightarrow$ Invest ~& 0.59\\ \hline
Asset\_Amt $\rightarrow$ Invest ~& 0.4\\ \hline
Asset\_Amt $\rightarrow$ Myopic\_Bias ~& 0.02\\ \hline
Asset\_Amt $\leftrightarrow$ Planning ~& 0.21\\ \hline
Asset\_Amt $\rightarrow$ Planning ~& 0.16\\ \hline
Asset\_Amt $\circ$–$\circ$ Planning ~& 0.01\\ \hline
Confidence $\leftrightarrow$ Asset\_Amt ~& 0.52\\ \hline
Confidence $\rightarrow$ Asset\_Amt ~& 0.14\\ \hline
Confidence $\circ$$\rightarrow$  Asset\_Amt ~& 0.04\\ \hline
Confidence $\leftrightarrow$ Fin\_Literacy ~& 0.09\\ \hline
Confidence $\leftrightarrow$ Income ~& 0.17\\ \hline
Confidence $\rightarrow$ Income ~& 0.06\\ \hline
Confidence $\circ$–$\circ$ Income ~& 0.01\\ \hline
Confidence $\circ$$\rightarrow$  Income ~& 0.01\\ \hline
Confidence $\rightarrow$ Invest ~& 0.35\\ \hline
Confidence $\leftrightarrow$ Invest ~& 0.11\\ \hline
Confidence $\circ$$\rightarrow$  Invest ~& 0.05\\ \hline
Confidence $\circ$–$\circ$ Invest ~& 0.01\\ \hline
Confidence $\leftrightarrow$ Planning ~& 0.41\\ \hline
Confidence $\rightarrow$ Planning ~& 0.31\\ \hline
Confidence $\circ$$\rightarrow$  Planning ~& 0.05\\ \hline
Education $\rightarrow$ Asset\_Amt ~& 0.45\\ \hline
Education $\rightarrow$ Confidence ~& 0.18\\ \hline
Education $\rightarrow$ Fin\_Literacy ~& 1\\ \hline
Education $\rightarrow$ Herding\_Bias ~& 0.05\\ \hline
Education $\rightarrow$ Income ~& 0.99\\ \hline
Education $\rightarrow$ Invest ~& 0.95\\ \hline
Education $\rightarrow$ Myopic\_Bias ~& 0.16\\ \hline
Education $\rightarrow$ Planning ~& 0.04\\ \hline
Fin\_Literacy $\rightarrow$ Asset\_Amt ~& 0.64\\ \hline
Fin\_Literacy $\circ$$\rightarrow$  Asset\_Amt ~& 0.05\\ \hline
Fin\_Literacy $\leftrightarrow$ Asset\_Amt ~& 0.05\\ \hline
Fin\_Literacy $\rightarrow$ Herding\_Bias ~& 0.62\\ \hline
Fin\_Literacy $\rightarrow$ Myopic\_Bias ~& 0.3\\ \hline
Herding\_Bias $\leftrightarrow$ Confidence ~& 0.27\\ \hline
Herding\_Bias $\leftrightarrow$ Fin\_Literacy ~& 0.37\\ \hline
Herding\_Bias $\circ$$\rightarrow$  Fin\_Literacy ~& 0.01\\ \hline
Herding\_Bias $\leftrightarrow$ Invest ~& 0.99\\ \hline
Herding\_Bias $\leftrightarrow$ Myopic\_Bias ~& 0.12\\ \hline
Herding\_Bias $\leftrightarrow$ Planning ~& 0.01\\ \hline
Income $\rightarrow$ Asset\_Amt ~& 0.72\\ \hline
Income $\circ$$\rightarrow$  Asset\_Amt ~& 0.05\\ \hline
Income $\leftrightarrow$ Asset\_Amt ~& 0.01\\ \hline
Income $\rightarrow$ Confidence ~& 0.26\\ \hline
Income $\leftrightarrow$ Invest ~& 0.22\\ \hline
Income $\rightarrow$ Invest ~& 0.15\\ \hline
Income $\circ$$\rightarrow$  Invest ~& 0.08\\ \hline
Income $\rightarrow$ Myopic\_Bias ~& 0.01\\ \hline
Income $\rightarrow$ Planning ~& 0.44\\ \hline
Income $\leftrightarrow$ Planning ~& 0.39\\ \hline
Income $\circ$$\rightarrow$  Planning ~& 0.01\\ \hline
Invest $\rightarrow$ Asset\_Amt ~& 0.01\\ \hline
Invest $\rightarrow$ Confidence ~& 0.48\\ \hline
Invest $\rightarrow$ Planning ~& 0.04\\ \hline
Myopic\_Bias $\leftrightarrow$ Asset\_Amt ~& 0.25\\ \hline
Myopic\_Bias $\circ$$\rightarrow$  Asset\_Amt ~& 0.14\\ \hline
Myopic\_Bias $\rightarrow$ Asset\_Amt ~& 0.13\\ \hline
Myopic\_Bias $\leftrightarrow$ Fin\_Literacy ~& 0.35\\ \hline
Myopic\_Bias $\circ$$\rightarrow$  Fin\_Literacy ~& 0.3\\ \hline
Myopic\_Bias $\rightarrow$ Fin\_Literacy ~& 0.03\\ \hline
Myopic\_Bias $\circ$–$\circ$ Fin\_Literacy ~& 0.02\\ \hline
Myopic\_Bias $\rightarrow$ Herding\_Bias ~& 0.05\\ \hline
Myopic\_Bias $\circ$$\rightarrow$  Herding\_Bias ~& 0.02\\ \hline
Myopic\_Bias $\leftrightarrow$ Income ~& 0.35\\ \hline
Myopic\_Bias $\circ$$\rightarrow$  Income ~& 0.04\\ \hline
Myopic\_Bias $\rightarrow$ Income ~& 0.02\\ \hline
Planning $\circ$$\rightarrow$  Asset\_Amt ~& 0.06\\ \hline
Planning $\rightarrow$ Asset\_Amt ~& 0.02\\ \hline
Planning $\rightarrow$ Confidence ~& 0.22\\ \hline
Planning $\circ$$\rightarrow$  Confidence ~& 0.01\\ \hline
Planning $\rightarrow$ Income ~& 0.12\\ \hline
Planning $\circ$$\rightarrow$  Income ~& 0.02\\ \hline
Planning $\leftrightarrow$ Invest ~& 0.37\\ \hline
Planning $\rightarrow$ Invest ~& 0.03\\ \hline

\end{supertabular}

\vspace{20pt} 

\tablefirsthead{\hline
    \multicolumn{2}{|c|}{\textbf{Group 3}} \\
    \hline
    \textbf{Edge} & \textbf{Prob.} \\ 
    \hline
}
\tablehead{\hline
    \multicolumn{2}{|c|}{\textbf{Group 3 (Continued)}} \\
    \hline
    \textbf{Edge} & \textbf{Prob.} \\
    \hline
}
\tabletail{
    \hline
    \multicolumn{2}{|r|}{\textit{To be Continued}} \\
    \hline
}
\tablelasttail{\hline}
\begin{supertabular}{|p{5cm}|p{1cm}|}
Age $\rightarrow$ Asset\_Amt ~& 1\\ \hline
Age $\rightarrow$ Fin\_Literacy ~& 1\\ \hline
Age $\rightarrow$ Herding\_Bias ~& 0.45\\ \hline
Age $\rightarrow$ Myopic\_Bias ~& 0.09\\ \hline
Age $\rightarrow$ Planning ~& 0.98\\ \hline
Asset\_Amt $\rightarrow$ Confidence ~& 0.06\\ \hline
Asset\_Amt $\rightarrow$ Fin\_Literacy ~& 0.62\\ \hline
Asset\_Amt $\rightarrow$ Herding\_Bias ~& 0.02\\ \hline
Asset\_Amt $\rightarrow$ Income ~& 0.03\\ \hline
Asset\_Amt $\leftrightarrow$ Invest ~& 0.1\\ \hline
Asset\_Amt $\rightarrow$ Planning ~& 0.13\\ \hline
Asset\_Amt $\leftrightarrow$ Planning ~& 0.13\\ \hline
Confidence $\circ$$\rightarrow$  Asset\_Amt ~& 0.31\\ \hline
Confidence $\leftrightarrow$ Asset\_Amt ~& 0.2\\ \hline
Confidence $\rightarrow$ Asset\_Amt ~& 0.13\\ \hline
Confidence $\leftrightarrow$ Fin\_Literacy ~& 0.08\\ \hline
Confidence $\circ$$\rightarrow$  Fin\_Literacy ~& 0.04\\ \hline
Confidence $\rightarrow$ Fin\_Literacy ~& 0.02\\ \hline
Confidence $\circ$$\rightarrow$  Income ~& 0.03\\ \hline
Confidence $\circ$–$\circ$ Income ~& 0.02\\ \hline
Confidence $\leftrightarrow$ Invest ~& 0.37\\ \hline
Confidence $\circ$$\rightarrow$  Invest ~& 0.36\\ \hline
Confidence $\rightarrow$ Invest ~& 0.19\\ \hline
Confidence $\circ$–$\circ$ Invest ~& 0.02\\ \hline
Confidence $\leftrightarrow$ Planning ~& 0.07\\ \hline
Confidence $\rightarrow$ Planning ~& 0.02\\ \hline
Confidence $\circ$$\rightarrow$  Planning ~& 0.01\\ \hline
Education $\rightarrow$ Asset\_Amt ~& 0.03\\ \hline
Education $\rightarrow$ Confidence ~& 0.41\\ \hline
Education $\rightarrow$ Fin\_Literacy ~& 0.75\\ \hline
Education $\rightarrow$ Income ~& 0.89\\ \hline
Education $\rightarrow$ Invest ~& 0.05\\ \hline
Education $\rightarrow$ Myopic\_Bias ~& 0.05\\ \hline
Education $\rightarrow$ Planning ~& 0.04\\ \hline
Fin\_Literacy $\rightarrow$ Asset\_Amt ~& 0.24\\ \hline
Fin\_Literacy $\leftrightarrow$ Asset\_Amt ~& 0.06\\ \hline
Fin\_Literacy $\rightarrow$ Confidence ~& 0.03\\ \hline
Fin\_Literacy $\rightarrow$ Herding\_Bias ~& 0.27\\ \hline
Fin\_Literacy $\leftrightarrow$ Invest ~& 0.72\\ \hline
Fin\_Literacy $\rightarrow$ Invest ~& 0.01\\ \hline
Fin\_Literacy $\leftrightarrow$ Planning ~& 0.5\\ \hline
Fin\_Literacy $\rightarrow$ Planning ~& 0.07\\ \hline
Herding\_Bias $\leftrightarrow$ Asset\_Amt ~& 0.11\\ \hline
Herding\_Bias $\circ$$\rightarrow$  Asset\_Amt ~& 0.07\\ \hline
Herding\_Bias $\circ$$\rightarrow$  Fin\_Literacy ~& 0.46\\ \hline
Herding\_Bias $\rightarrow$ Fin\_Literacy ~& 0.2\\ \hline
Herding\_Bias $\leftrightarrow$ Fin\_Literacy ~& 0.07\\ \hline
Herding\_Bias $\leftrightarrow$ Invest ~& 0.36\\ \hline
Herding\_Bias $\circ$$\rightarrow$  Invest ~& 0.11\\ \hline
Herding\_Bias $\rightarrow$ Invest ~& 0.1\\ \hline
Herding\_Bias $\rightarrow$ Myopic\_Bias ~& 0.03\\ \hline
Herding\_Bias $\leftrightarrow$ Myopic\_Bias ~& 0.02\\ \hline
Herding\_Bias $\circ$$\rightarrow$  Myopic\_Bias ~& 0.02\\ \hline
Herding\_Bias $\circ$–$\circ$ Myopic\_Bias ~& 0.02\\ \hline
Income $\rightarrow$ Asset\_Amt ~& 0.91\\ \hline
Income $\leftrightarrow$ Asset\_Amt ~& 0.05\\ \hline
Income $\circ$$\rightarrow$  Asset\_Amt ~& 0.01\\ \hline
Income $\rightarrow$ Confidence ~& 0.08\\ \hline
Income $\leftrightarrow$ Invest ~& 0.68\\ \hline
Income $\rightarrow$ Invest ~& 0.12\\ \hline
Income $\circ$$\rightarrow$  Invest ~& 0.02\\ \hline
Income $\leftrightarrow$ Planning ~& 0.67\\ \hline
Income $\rightarrow$ Planning ~& 0.16\\ \hline
Invest $\rightarrow$ Asset\_Amt ~& 0.06\\ \hline
Invest $\rightarrow$ Confidence ~& 0.06\\ \hline
Invest $\rightarrow$ Fin\_Literacy ~& 0.01\\ \hline
Invest $\circ$$\rightarrow$  Income ~& 0.02\\ \hline
Myopic\_Bias $\circ$$\rightarrow$  Asset\_Amt ~& 0.37\\ \hline
Myopic\_Bias $\leftrightarrow$ Asset\_Amt ~& 0.16\\ \hline
Myopic\_Bias $\rightarrow$ Asset\_Amt ~& 0.08\\ \hline
Myopic\_Bias $\leftrightarrow$ Confidence ~& 0.04\\ \hline
Myopic\_Bias $\circ$$\rightarrow$  Confidence ~& 0.01\\ \hline
Myopic\_Bias $\circ$$\rightarrow$  Fin\_Literacy ~& 0.29\\ \hline
Myopic\_Bias $\leftrightarrow$ Fin\_Literacy ~& 0.1\\ \hline
Myopic\_Bias $\rightarrow$ Fin\_Literacy ~& 0.07\\ \hline
Myopic\_Bias $\circ$$\rightarrow$  Herding\_Bias ~& 0.04\\ \hline
Myopic\_Bias $\leftrightarrow$ Income ~& 0.01\\ \hline
Myopic\_Bias $\leftrightarrow$ Invest ~& 0.02\\ \hline
Myopic\_Bias $\circ$$\rightarrow$  Invest ~& 0.01\\ \hline
Myopic\_Bias $\circ$$\rightarrow$  Planning ~& 0.04\\ \hline
Myopic\_Bias $\leftrightarrow$ Planning ~& 0.02\\ \hline
Planning $\rightarrow$ Asset\_Amt ~& 0.2\\ \hline
Planning $\circ$$\rightarrow$  Asset\_Amt ~& 0.02\\ \hline
Planning $\rightarrow$ Confidence ~& 0.01\\ \hline
Planning $\rightarrow$ Fin\_Literacy ~& 0.12\\ \hline
Planning $\circ$$\rightarrow$  Fin\_Literacy ~& 0.07\\ \hline
Planning $\rightarrow$ Income ~& 0.02\\ \hline
Planning $\leftrightarrow$ Invest ~& 0.01\\ \hline
Planning $\rightarrow$ Myopic\_Bias ~& 0.01\\ \hline
\end{supertabular}

\vspace{20pt} 

\tablefirsthead{\hline
    \multicolumn{2}{|c|}{\textbf{Group 4}} \\
    \hline
    \textbf{Edge} & \textbf{Prob.} \\ 
    \hline
}
\tablehead{\hline
    \multicolumn{2}{|c|}{\textbf{Group 4 (Continued)}} \\
    \hline
    \textbf{Edge} & \textbf{Prob.} \\
    \hline
}
\tabletail{
    \hline
    \multicolumn{2}{|r|}{\textit{To be Continued}} \\
    \hline
}
\tablelasttail{\hline}
\begin{supertabular}{|p{5cm}|p{1cm}|}
Age $\rightarrow$ Asset\_Amt ~& 1\\ \hline
Age $\rightarrow$ Fin\_Literacy ~& 1\\ \hline
Age $\rightarrow$ Herding\_Bias ~& 1\\ \hline
Age $\rightarrow$ Myopic\_Bias ~& 1\\ \hline
Age $\rightarrow$ Planning ~& 1\\ \hline
Asset\_Amt $\rightarrow$ Confidence ~& 0.23\\ \hline
Asset\_Amt $\circ$$\rightarrow$  Fin\_Literacy ~& 0.18\\ \hline
Asset\_Amt $\rightarrow$ Fin\_Literacy ~& 0.15\\ \hline
Asset\_Amt $\leftrightarrow$ Invest ~& 0.58\\ \hline
Asset\_Amt $\rightarrow$ Invest ~& 0.04\\ \hline
Asset\_Amt $\rightarrow$ Myopic\_Bias ~& 0.66\\ \hline
Asset\_Amt $\rightarrow$ Planning ~& 0.43\\ \hline
Asset\_Amt $\circ$$\rightarrow$  Planning ~& 0.02\\ \hline
Asset\_Amt $\leftrightarrow$ Planning ~& 0.01\\ \hline
Confidence $\leftrightarrow$ Asset\_Amt ~& 0.46\\ \hline
Confidence $\rightarrow$ Asset\_Amt ~& 0.31\\ \hline
Confidence $\circ$$\rightarrow$  Fin\_Literacy ~& 0.35\\ \hline
Confidence $\leftrightarrow$ Fin\_Literacy ~& 0.24\\ \hline
Confidence $\rightarrow$ Fin\_Literacy ~& 0.16\\ \hline
Confidence $\circ$–$\circ$ Income ~& 0.18\\ \hline
Confidence $\leftrightarrow$ Income ~& 0.18\\ \hline
Confidence $\rightarrow$ Income ~& 0.14\\ \hline
Confidence $\circ$$\rightarrow$  Income ~& 0.01\\ \hline
Confidence $\rightarrow$ Invest ~& 0.42\\ \hline
Confidence $\circ$–$\circ$ Invest ~& 0.14\\ \hline
Confidence $\circ$$\rightarrow$  Invest ~& 0.1\\ \hline
Confidence $\leftrightarrow$ Invest ~& 0.09\\ \hline
Confidence $\rightarrow$ Myopic\_Bias ~& 0.05\\ \hline
Confidence $\circ$$\rightarrow$  Myopic\_Bias ~& 0.01\\ \hline
Confidence $\leftrightarrow$ Planning ~& 0.48\\ \hline
Confidence $\rightarrow$ Planning ~& 0.44\\ \hline
Confidence $\circ$$\rightarrow$  Planning ~& 0.01\\ \hline
Education $\rightarrow$ Asset\_Amt ~& 0.99\\ \hline
Education $\rightarrow$ Confidence ~& 1\\ \hline
Education $\rightarrow$ Fin\_Literacy ~& 1\\ \hline
Education $\rightarrow$ Income ~& 1\\ \hline
Education $\rightarrow$ Invest ~& 1\\ \hline
Education $\rightarrow$ Myopic\_Bias ~& 1\\ \hline
Education $\rightarrow$ Planning ~& 0.11\\ \hline
Fin\_Literacy $\rightarrow$ Asset\_Amt ~& 0.59\\ \hline
Fin\_Literacy $\leftrightarrow$ Asset\_Amt ~& 0.08\\ \hline
Fin\_Literacy $\rightarrow$ Confidence ~& 0.25\\ \hline
Fin\_Literacy $\leftrightarrow$ Income ~& 0.34\\ \hline
Fin\_Literacy $\leftrightarrow$ Invest ~& 0.4\\ \hline
Fin\_Literacy $\rightarrow$ Invest ~& 0.07\\ \hline
Fin\_Literacy $\rightarrow$ Myopic\_Bias ~& 0.06\\ \hline
Fin\_Literacy $\leftrightarrow$ Planning ~& 0.42\\ \hline
Fin\_Literacy $\rightarrow$ Planning ~& 0.09\\ \hline
Herding\_Bias $\leftrightarrow$ Confidence ~& 0.36\\ \hline
Herding\_Bias $\rightarrow$ Confidence ~& 0.06\\ \hline
Herding\_Bias $\circ$$\rightarrow$  Fin\_Literacy ~& 0.65\\ \hline
Herding\_Bias $\leftrightarrow$ Fin\_Literacy ~& 0.31\\ \hline
Herding\_Bias $\rightarrow$ Fin\_Literacy ~& 0.04\\ \hline
Herding\_Bias $\leftrightarrow$ Income ~& 0.19\\ \hline
Herding\_Bias $\leftrightarrow$ Invest ~& 0.08\\ \hline
Income $\rightarrow$ Asset\_Amt ~& 0.45\\ \hline
Income $\leftrightarrow$ Asset\_Amt ~& 0.38\\ \hline
Income $\circ$$\rightarrow$  Asset\_Amt ~& 0.17\\ \hline
Income $\rightarrow$ Confidence ~& 0.34\\ \hline
Income $\circ$$\rightarrow$  Confidence ~& 0.15\\ \hline
Income $\circ$$\rightarrow$  Fin\_Literacy ~& 0.5\\ \hline
Income $\rightarrow$ Fin\_Literacy ~& 0.11\\ \hline
Income $\leftrightarrow$ Invest ~& 0.15\\ \hline
Income $\rightarrow$ Invest ~& 0.15\\ \hline
Income $\circ$–$\circ$ Invest ~& 0.12\\ \hline
Income $\circ$$\rightarrow$  Invest ~& 0.08\\ \hline
Income $\circ$$\rightarrow$  Myopic\_Bias ~& 0.01\\ \hline
Income $\rightarrow$ Planning ~& 0.55\\ \hline
Income $\leftrightarrow$ Planning ~& 0.34\\ \hline
Income $\circ$$\rightarrow$  Planning ~& 0.03\\ \hline
Invest $\rightarrow$ Asset\_Amt ~& 0.28\\ \hline
Invest $\circ$$\rightarrow$  Asset\_Amt ~& 0.1\\ \hline
Invest $\rightarrow$ Confidence ~& 0.18\\ \hline
Invest $\circ$$\rightarrow$  Confidence ~& 0.07\\ \hline
Invest $\circ$$\rightarrow$  Fin\_Literacy ~& 0.41\\ \hline
Invest $\rightarrow$ Fin\_Literacy ~& 0.12\\ \hline
Invest $\circ$$\rightarrow$  Income ~& 0.07\\ \hline
Invest $\rightarrow$ Income ~& 0.04\\ \hline
Invest $\circ$$\rightarrow$  Myopic\_Bias ~& 0.01\\ \hline
Invest $\rightarrow$ Planning ~& 0.31\\ \hline
Invest $\circ$$\rightarrow$  Planning ~& 0.01\\ \hline
Myopic\_Bias $\rightarrow$ Asset\_Amt ~& 0.34\\ \hline
Myopic\_Bias $\leftrightarrow$ Confidence ~& 0.25\\ \hline
Myopic\_Bias $\rightarrow$ Confidence ~& 0.06\\ \hline
Myopic\_Bias $\circ$$\rightarrow$  Fin\_Literacy ~& 0.4\\ \hline
Myopic\_Bias $\leftrightarrow$ Fin\_Literacy ~& 0.09\\ \hline
Myopic\_Bias $\leftrightarrow$ Income ~& 0.11\\ \hline
Myopic\_Bias $\leftrightarrow$ Invest ~& 0.01\\ \hline
Myopic\_Bias $\leftrightarrow$ Planning ~& 0.09\\ \hline
Myopic\_Bias $\rightarrow$ Planning ~& 0.04\\ \hline
Planning $\rightarrow$ Asset\_Amt ~& 0.51\\ \hline
Planning $\circ$$\rightarrow$  Asset\_Amt ~& 0.03\\ \hline
Planning $\rightarrow$ Confidence ~& 0.07\\ \hline
Planning $\circ$$\rightarrow$  Fin\_Literacy ~& 0.35\\ \hline
Planning $\rightarrow$ Fin\_Literacy ~& 0.14\\ \hline
Planning $\rightarrow$ Income ~& 0.08\\ \hline
Planning $\leftrightarrow$ Invest ~& 0.54\\ \hline
Planning $\rightarrow$ Invest ~& 0.03\\ \hline
Planning $\rightarrow$ Myopic\_Bias ~& 0.03\\ \hline

\end{supertabular}

\vspace{20pt} 

\tablefirsthead{\hline
    \multicolumn{2}{|c|}{\textbf{Group 5}} \\
    \hline
    \textbf{Edge} & \textbf{Prob.} \\ 
    \hline
}
\tablehead{\hline
    \multicolumn{2}{|c|}{\textbf{Group 5 (Continued)}} \\
    \hline
    \textbf{Edge} & \textbf{Prob.} \\
    \hline
}
\tabletail{
    \hline
    \multicolumn{2}{|r|}{\textit{To be Continued}} \\
    \hline
}
\tablelasttail{\hline}
\begin{supertabular}{|p{5cm}|p{1cm}|}
Age $\rightarrow$ Asset\_Amt ~& 1\\ \hline
Age $\rightarrow$ Confidence ~& 0.01\\ \hline
Age $\rightarrow$ Fin\_Literacy ~& 0.19\\ \hline
Age $\rightarrow$ Income ~& 0.02\\ \hline
Age $\rightarrow$ Invest ~& 0.33\\ \hline
Age $\rightarrow$ Myopic\_Bias ~& 0.14\\ \hline
Age $\rightarrow$ Planning ~& 0.49\\ \hline
Asset\_Amt $\rightarrow$ Confidence ~& 0.08\\ \hline
Asset\_Amt $\rightarrow$ Fin\_Literacy ~& 0.19\\ \hline
Asset\_Amt $\circ$$\rightarrow$  Fin\_Literacy ~& 0.01\\ \hline
Asset\_Amt $\rightarrow$ Income ~& 0.62\\ \hline
Asset\_Amt $\rightarrow$ Invest ~& 0.63\\ \hline
Asset\_Amt $\leftrightarrow$ Invest ~& 0.13\\ \hline
Asset\_Amt $\circ$$\rightarrow$  Invest ~& 0.08\\ \hline
Asset\_Amt $\circ$–$\circ$ Invest ~& 0.05\\ \hline
Confidence $\rightarrow$ Asset\_Amt ~& 0.02\\ \hline
Confidence $\leftrightarrow$ Asset\_Amt ~& 0.02\\ \hline
Confidence $\rightarrow$ Fin\_Literacy ~& 0.13\\ \hline
Confidence $\leftrightarrow$ Fin\_Literacy ~& 0.08\\ \hline
Confidence $\circ$$\rightarrow$  Fin\_Literacy ~& 0.07\\ \hline
Confidence $\rightarrow$ Herding\_Bias ~& 0.06\\ \hline
Confidence $\circ$$\rightarrow$  Herding\_Bias ~& 0.03\\ \hline
Confidence $\leftrightarrow$ Income ~& 0.01\\ \hline
Confidence $\circ$$\rightarrow$  Invest ~& 0.16\\ \hline
Confidence $\leftrightarrow$ Invest ~& 0.07\\ \hline
Confidence $\rightarrow$ Invest ~& 0.07\\ \hline
Confidence $\circ$–$\circ$ Invest ~& 0.01\\ \hline
Confidence $\rightarrow$ Planning ~& 0.07\\ \hline
Confidence $\leftrightarrow$ Planning ~& 0.07\\ \hline
Confidence $\circ$$\rightarrow$  Planning ~& 0.01\\ \hline
Education $\rightarrow$ Confidence ~& 0.04\\ \hline
Education $\rightarrow$ Fin\_Literacy ~& 0.04\\ \hline
Education $\rightarrow$ Herding\_Bias ~& 0.03\\ \hline
Education $\rightarrow$ Income ~& 0.63\\ \hline
Education $\rightarrow$ Myopic\_Bias ~& 0.12\\ \hline
Fin\_Literacy $\leftrightarrow$ Asset\_Amt ~& 0.3\\ \hline
Fin\_Literacy $\rightarrow$ Asset\_Amt ~& 0.01\\ \hline
Fin\_Literacy $\circ$$\rightarrow$  Asset\_Amt ~& 0.01\\ \hline
Fin\_Literacy $\rightarrow$ Confidence ~& 0.02\\ \hline
Fin\_Literacy $\rightarrow$ Myopic\_Bias ~& 0.01\\ \hline
Fin\_Literacy $\leftrightarrow$ Planning ~& 0.29\\ \hline
Herding\_Bias $\leftrightarrow$ Confidence ~& 0.18\\ \hline
Herding\_Bias $\rightarrow$ Confidence ~& 0.01\\ \hline
Herding\_Bias $\leftrightarrow$ Fin\_Literacy ~& 0.82\\ \hline
Herding\_Bias $\circ$$\rightarrow$  Fin\_Literacy ~& 0.13\\ \hline
Herding\_Bias $\leftrightarrow$ Income ~& 0.05\\ \hline
Herding\_Bias $\leftrightarrow$ Invest ~& 0.67\\ \hline
Herding\_Bias $\circ$$\rightarrow$  Invest ~& 0.04\\ \hline
Income $\leftrightarrow$ Asset\_Amt ~& 0.25\\ \hline
Income $\rightarrow$ Asset\_Amt ~& 0.07\\ \hline
Income $\circ$$\rightarrow$  Asset\_Amt ~& 0.06\\ \hline
Income $\leftrightarrow$ Invest ~& 0.02\\ \hline
Income $\rightarrow$ Invest ~& 0.02\\ \hline
Income $\rightarrow$ Myopic\_Bias ~& 0.02\\ \hline
Income $\leftrightarrow$ Planning ~& 0.3\\ \hline
Income $\rightarrow$ Planning ~& 0.21\\ \hline
Income $\circ$$\rightarrow$  Planning ~& 0.03\\ \hline
Invest $\rightarrow$ Asset\_Amt ~& 0.08\\ \hline
Invest $\circ$$\rightarrow$  Asset\_Amt ~& 0.01\\ \hline
Invest $\rightarrow$ Confidence ~& 0.68\\ \hline
Invest $\circ$$\rightarrow$  Confidence ~& 0.01\\ \hline
Invest $\rightarrow$ Fin\_Literacy ~& 0.02\\ \hline
Invest $\rightarrow$ Herding\_Bias ~& 0.02\\ \hline
Invest $\circ$$\rightarrow$  Income ~& 0.01\\ \hline
Invest $\rightarrow$ Income ~& 0.01\\ \hline
Myopic\_Bias $\leftrightarrow$ Confidence ~& 0.02\\ \hline
Myopic\_Bias $\circ$$\rightarrow$  Confidence ~& 0.01\\ \hline
Myopic\_Bias $\circ$$\rightarrow$  Fin\_Literacy ~& 0.63\\ \hline
Myopic\_Bias $\leftrightarrow$ Fin\_Literacy ~& 0.32\\ \hline
Myopic\_Bias $\rightarrow$ Fin\_Literacy ~& 0.02\\ \hline
Myopic\_Bias $\circ$–$\circ$ Fin\_Literacy ~& 0.01\\ \hline
Myopic\_Bias $\circ$$\rightarrow$  Herding\_Bias ~& 0.01\\ \hline
Myopic\_Bias $\leftrightarrow$ Income ~& 0.09\\ \hline
Myopic\_Bias $\circ$$\rightarrow$  Income ~& 0.03\\ \hline
Myopic\_Bias $\leftrightarrow$ Invest ~& 0.03\\ \hline
Myopic\_Bias $\circ$$\rightarrow$  Invest ~& 0.02\\ \hline
Myopic\_Bias $\leftrightarrow$ Planning ~& 0.02\\ \hline
Myopic\_Bias $\circ$$\rightarrow$  Planning ~& 0.01\\ \hline
Planning $\circ$$\rightarrow$  Confidence ~& 0.02\\ \hline
Planning $\rightarrow$ Confidence ~& 0.01\\ \hline
Planning $\rightarrow$ Fin\_Literacy ~& 0.1\\ \hline
Planning $\circ$$\rightarrow$  Fin\_Literacy ~& 0.07\\ \hline
Planning $\circ$$\rightarrow$  Income ~& 0.17\\ \hline

\end{supertabular}

\vspace{20pt} 

\tablefirsthead{\hline
    \multicolumn{2}{|c|}{\textbf{Group 6}} \\
    \hline
    \textbf{Edge} & \textbf{Prob.} \\ 
    \hline
}
\tablehead{\hline
    \multicolumn{2}{|c|}{\textbf{Group 6 (Continued)}} \\
    \hline
    \textbf{Edge} & \textbf{Prob.} \\
    \hline
}
\tabletail{
    \hline
    \multicolumn{2}{|r|}{\textit{To be Continued}} \\
    \hline
}
\tablelasttail{\hline}
\begin{supertabular}{|p{5cm}|p{1cm}|}
Age $\rightarrow$ Asset\_Amt ~& 1\\ \hline
Age $\rightarrow$ Confidence ~& 0.01\\ \hline
Age $\rightarrow$ Fin\_Literacy ~& 0.97\\ \hline
Age $\rightarrow$ Herding\_Bias ~& 0.1\\ \hline
Age $\rightarrow$ Income ~& 0.02\\ \hline
Age $\rightarrow$ Invest ~& 0.95\\ \hline
Age $\rightarrow$ Myopic\_Bias ~& 0.97\\ \hline
Asset\_Amt $\rightarrow$ Confidence ~& 0.52\\ \hline
Asset\_Amt $\rightarrow$ Fin\_Literacy ~& 0.46\\ \hline
Asset\_Amt $\circ$$\rightarrow$  Fin\_Literacy ~& 0.26\\ \hline
Asset\_Amt $\rightarrow$ Invest ~& 0.72\\ \hline
Asset\_Amt $\leftrightarrow$ Invest ~& 0.04\\ \hline
Asset\_Amt $\rightarrow$ Myopic\_Bias ~& 0.08\\ \hline
Asset\_Amt $\rightarrow$ Planning ~& 0.05\\ \hline
Confidence $\rightarrow$ Asset\_Amt ~& 0.22\\ \hline
Confidence $\leftrightarrow$ Asset\_Amt ~& 0.19\\ \hline
Confidence $\circ$$\rightarrow$  Asset\_Amt ~& 0.05\\ \hline
Confidence $\circ$$\rightarrow$  Fin\_Literacy ~& 0.43\\ \hline
Confidence $\leftrightarrow$ Fin\_Literacy ~& 0.3\\ \hline
Confidence $\rightarrow$ Fin\_Literacy ~& 0.16\\ \hline
Confidence $\leftrightarrow$ Income ~& 0.08\\ \hline
Confidence $\rightarrow$ Invest ~& 0.4\\ \hline
Confidence $\leftrightarrow$ Invest ~& 0.14\\ \hline
Confidence $\circ$$\rightarrow$  Invest ~& 0.09\\ \hline
Confidence $\circ$–$\circ$ Invest ~& 0.01\\ \hline
Confidence $\rightarrow$ Planning ~& 0.38\\ \hline
Confidence $\leftrightarrow$ Planning ~& 0.26\\ \hline
Confidence $\circ$–$\circ$ Planning ~& 0.02\\ \hline
Confidence $\circ$$\rightarrow$  Planning ~& 0.01\\ \hline
Education $\rightarrow$ Confidence ~& 0.15\\ \hline
Education $\rightarrow$ Fin\_Literacy ~& 0.99\\ \hline
Education $\rightarrow$ Herding\_Bias ~& 0.06\\ \hline
Education $\rightarrow$ Income ~& 0.96\\ \hline
Education $\rightarrow$ Invest ~& 0.59\\ \hline
Education $\rightarrow$ Myopic\_Bias ~& 0.16\\ \hline
Education $\rightarrow$ Planning ~& 0.13\\ \hline
Fin\_Literacy $\rightarrow$ Asset\_Amt ~& 0.25\\ \hline
Fin\_Literacy $\rightarrow$ Confidence ~& 0.01\\ \hline
Fin\_Literacy $\leftrightarrow$ Income ~& 0.02\\ \hline
Fin\_Literacy $\leftrightarrow$ Invest ~& 0.17\\ \hline
Fin\_Literacy $\rightarrow$ Invest ~& 0.02\\ \hline
Fin\_Literacy $\rightarrow$ Myopic\_Bias ~& 0.13\\ \hline
Fin\_Literacy $\rightarrow$ Planning ~& 0.15\\ \hline
Fin\_Literacy $\leftrightarrow$ Planning ~& 0.09\\ \hline
Herding\_Bias $\circ$$\rightarrow$  Fin\_Literacy ~& 0.94\\ \hline
Herding\_Bias $\leftrightarrow$ Fin\_Literacy ~& 0.06\\ \hline
Herding\_Bias $\leftrightarrow$ Income ~& 0.02\\ \hline
Herding\_Bias $\leftrightarrow$ Invest ~& 0.03\\ \hline
Herding\_Bias $\circ$$\rightarrow$  Invest ~& 0.02\\ \hline
Herding\_Bias $\rightarrow$ Invest ~& 0.01\\ \hline
Herding\_Bias $\circ$$\rightarrow$  Myopic\_Bias ~& 0.06\\ \hline
Herding\_Bias $\leftrightarrow$ Myopic\_Bias ~& 0.02\\ \hline
Herding\_Bias $\rightarrow$ Myopic\_Bias ~& 0.01\\ \hline
Herding\_Bias $\leftrightarrow$ Planning ~& 0.01\\ \hline
Income $\rightarrow$ Asset\_Amt ~& 0.91\\ \hline
Income $\leftrightarrow$ Asset\_Amt ~& 0.08\\ \hline
Income $\circ$$\rightarrow$  Asset\_Amt ~& 0.01\\ \hline
Income $\rightarrow$ Confidence ~& 0.09\\ \hline
Income $\circ$$\rightarrow$  Fin\_Literacy ~& 0.4\\ \hline
Income $\rightarrow$ Fin\_Literacy ~& 0.03\\ \hline
Income $\rightarrow$ Invest ~& 0.02\\ \hline
Income $\circ$$\rightarrow$  Myopic\_Bias ~& 0.02\\ \hline
Income $\rightarrow$ Myopic\_Bias ~& 0.01\\ \hline
Income $\rightarrow$ Planning ~& 0.69\\ \hline
Income $\leftrightarrow$ Planning ~& 0.09\\ \hline
Income $\circ$–$\circ$ Planning ~& 0.05\\ \hline
Income $\circ$$\rightarrow$  Planning ~& 0.05\\ \hline
Invest $\rightarrow$ Asset\_Amt ~& 0.24\\ \hline
Invest $\rightarrow$ Confidence ~& 0.35\\ \hline
Invest $\circ$$\rightarrow$  Confidence ~& 0.01\\ \hline
Invest $\circ$$\rightarrow$  Fin\_Literacy ~& 0.66\\ \hline
Invest $\rightarrow$ Fin\_Literacy ~& 0.12\\ \hline
Invest $\rightarrow$ Planning ~& 0.09\\ \hline
Myopic\_Bias $\circ$$\rightarrow$  Asset\_Amt ~& 0.09\\ \hline
Myopic\_Bias $\leftrightarrow$ Asset\_Amt ~& 0.08\\ \hline
Myopic\_Bias $\rightarrow$ Asset\_Amt ~& 0.08\\ \hline
Myopic\_Bias $\circ$$\rightarrow$  Fin\_Literacy ~& 0.6\\ \hline
Myopic\_Bias $\leftrightarrow$ Fin\_Literacy ~& 0.16\\ \hline
Myopic\_Bias $\rightarrow$ Fin\_Literacy ~& 0.11\\ \hline
Myopic\_Bias $\leftrightarrow$ Income ~& 0.12\\ \hline
Myopic\_Bias $\leftrightarrow$ Planning ~& 0.12\\ \hline
Planning $\rightarrow$ Confidence ~& 0.28\\ \hline
Planning $\circ$$\rightarrow$  Confidence ~& 0.02\\ \hline
Planning $\circ$$\rightarrow$  Fin\_Literacy ~& 0.38\\ \hline
Planning $\rightarrow$ Fin\_Literacy ~& 0.37\\ \hline
Planning $\rightarrow$ Income ~& 0.08\\ \hline
Planning $\circ$$\rightarrow$  Income ~& 0.04\\ \hline
Planning $\leftrightarrow$ Invest ~& 0.05\\ \hline
Planning $\rightarrow$ Invest ~& 0.02\\ \hline
Planning $\rightarrow$ Myopic\_Bias ~& 0.01\\ \hline

\end{supertabular}

\vspace{20pt} 

\tablefirsthead{\hline
    \multicolumn{2}{|c|}{\textbf{Group 7}} \\
    \hline
    \textbf{Edge} & \textbf{Prob.} \\ 
    \hline
}
\tablehead{\hline
    \multicolumn{2}{|c|}{\textbf{Group 7 (Continued)}} \\
    \hline
    \textbf{Edge} & \textbf{Prob.} \\
    \hline
}
\tabletail{
    \hline
    \multicolumn{2}{|r|}{\textit{To be Continued}} \\
    \hline
}
\tablelasttail{\hline}
\begin{supertabular}{|p{5cm}|p{1cm}|}
Age $\rightarrow$ Asset\_Amt ~& 1\\ \hline
Age $\rightarrow$ Confidence ~& 0.17\\ \hline
Age $\rightarrow$ Fin\_Literacy ~& 0.24\\ \hline
Age $\rightarrow$ Herding\_Bias ~& 0.08\\ \hline
Age $\rightarrow$ Invest ~& 0.12\\ \hline
Age $\rightarrow$ Planning ~& 0.33\\ \hline
Asset\_Amt $\rightarrow$ Confidence ~& 0.09\\ \hline
Asset\_Amt $\rightarrow$ Fin\_Literacy ~& 0.19\\ \hline
Asset\_Amt $\rightarrow$ Income ~& 0.34\\ \hline
Asset\_Amt $\rightarrow$ Invest ~& 0.24\\ \hline
Asset\_Amt $\leftrightarrow$ Invest ~& 0.08\\ \hline
Asset\_Amt $\rightarrow$ Planning ~& 0.33\\ \hline
Asset\_Amt $\circ$$\rightarrow$  Planning ~& 0.02\\ \hline
Asset\_Amt $\circ$–$\circ$ Planning ~& 0.01\\ \hline
Asset\_Amt $\leftrightarrow$ Planning ~& 0.01\\ \hline
Confidence $\circ$$\rightarrow$  Asset\_Amt ~& 0.03\\ \hline
Confidence $\rightarrow$ Asset\_Amt ~& 0.03\\ \hline
Confidence $\leftrightarrow$ Fin\_Literacy ~& 0.3\\ \hline
Confidence $\rightarrow$ Fin\_Literacy ~& 0.01\\ \hline
Confidence $\rightarrow$ Herding\_Bias ~& 0.01\\ \hline
Confidence $\leftrightarrow$ Income ~& 0.1\\ \hline
Confidence $\circ$$\rightarrow$  Income ~& 0.04\\ \hline
Confidence $\rightarrow$ Income ~& 0.01\\ \hline
Confidence $\circ$–$\circ$ Invest ~& 0.2\\ \hline
Confidence $\rightarrow$ Invest ~& 0.16\\ \hline
Confidence $\leftrightarrow$ Invest ~& 0.15\\ \hline
Confidence $\circ$$\rightarrow$  Invest ~& 0.11\\ \hline
Confidence $\circ$$\rightarrow$  Planning ~& 0.01\\ \hline
Education $\rightarrow$ Asset\_Amt ~& 0.01\\ \hline
Education $\rightarrow$ Confidence ~& 0.01\\ \hline
Education $\rightarrow$ Fin\_Literacy ~& 0.2\\ \hline
Education $\rightarrow$ Herding\_Bias ~& 0.06\\ \hline
Education $\rightarrow$ Income ~& 0.78\\ \hline
Education $\rightarrow$ Myopic\_Bias ~& 0.64\\ \hline
Education $\rightarrow$ Planning ~& 0.12\\ \hline
Fin\_Literacy $\leftrightarrow$ Asset\_Amt ~& 0.3\\ \hline
Fin\_Literacy $\rightarrow$ Asset\_Amt ~& 0.1\\ \hline
Fin\_Literacy $\circ$$\rightarrow$  Asset\_Amt ~& 0.02\\ \hline
Fin\_Literacy $\leftrightarrow$ Invest ~& 0.23\\ \hline
Fin\_Literacy $\rightarrow$ Myopic\_Bias ~& 0.04\\ \hline
Fin\_Literacy $\circ$$\rightarrow$  Myopic\_Bias ~& 0.01\\ \hline
Fin\_Literacy $\leftrightarrow$ Planning ~& 0.25\\ \hline
Fin\_Literacy $\rightarrow$ Planning ~& 0.13\\ \hline
Fin\_Literacy $\circ$$\rightarrow$  Planning ~& 0.03\\ \hline
Herding\_Bias $\leftrightarrow$ Confidence ~& 0.03\\ \hline
Herding\_Bias $\circ$$\rightarrow$  Confidence ~& 0.01\\ \hline
Herding\_Bias $\circ$$\rightarrow$  Fin\_Literacy ~& 0.49\\ \hline
Herding\_Bias $\leftrightarrow$ Fin\_Literacy ~& 0.13\\ \hline
Herding\_Bias $\rightarrow$ Fin\_Literacy ~& 0.05\\ \hline
Herding\_Bias $\circ$–$\circ$ Fin\_Literacy ~& 0.02\\ \hline
Herding\_Bias $\leftrightarrow$ Income ~& 0.01\\ \hline
Herding\_Bias $\leftrightarrow$ Invest ~& 0.04\\ \hline
Herding\_Bias $\circ$$\rightarrow$  Invest ~& 0.01\\ \hline
Herding\_Bias $\leftrightarrow$ Myopic\_Bias ~& 0.02\\ \hline
Herding\_Bias $\circ$$\rightarrow$  Myopic\_Bias ~& 0.01\\ \hline
Herding\_Bias $\circ$–$\circ$ Planning ~& 0.01\\ \hline
Income $\rightarrow$ Asset\_Amt ~& 0.6\\ \hline
Income $\leftrightarrow$ Asset\_Amt ~& 0.03\\ \hline
Income $\circ$$\rightarrow$  Asset\_Amt ~& 0.03\\ \hline
Income $\rightarrow$ Confidence ~& 0.01\\ \hline
Income $\leftrightarrow$ Invest ~& 0.02\\ \hline
Income $\rightarrow$ Myopic\_Bias ~& 0.2\\ \hline
Income $\circ$$\rightarrow$  Myopic\_Bias ~& 0.05\\ \hline
Income $\rightarrow$ Planning ~& 0.26\\ \hline
Income $\leftrightarrow$ Planning ~& 0.07\\ \hline
Income $\circ$$\rightarrow$  Planning ~& 0.01\\ \hline
Invest $\circ$$\rightarrow$  Asset\_Amt ~& 0.28\\ \hline
Invest $\rightarrow$ Asset\_Amt ~& 0.09\\ \hline
Invest $\rightarrow$ Confidence ~& 0.2\\ \hline
Invest $\circ$$\rightarrow$  Confidence ~& 0.18\\ \hline
Invest $\circ$$\rightarrow$  Herding\_Bias ~& 0.01\\ \hline
Invest $\circ$$\rightarrow$  Income ~& 0.03\\ \hline
Invest $\rightarrow$ Income ~& 0.01\\ \hline
Invest $\circ$$\rightarrow$  Myopic\_Bias ~& 0.01\\ \hline
Invest $\rightarrow$ Planning ~& 0.02\\ \hline
Myopic\_Bias $\leftrightarrow$ Asset\_Amt ~& 0.03\\ \hline
Myopic\_Bias $\circ$$\rightarrow$  Asset\_Amt ~& 0.02\\ \hline
Myopic\_Bias $\rightarrow$ Confidence ~& 0.01\\ \hline
Myopic\_Bias $\leftrightarrow$ Confidence ~& 0.01\\ \hline
Myopic\_Bias $\rightarrow$ Fin\_Literacy ~& 0.36\\ \hline
Myopic\_Bias $\leftrightarrow$ Fin\_Literacy ~& 0.32\\ \hline
Myopic\_Bias $\circ$$\rightarrow$  Fin\_Literacy ~& 0.23\\ \hline
Myopic\_Bias $\circ$–$\circ$ Fin\_Literacy ~& 0.02\\ \hline
Myopic\_Bias $\leftrightarrow$ Income ~& 0.11\\ \hline
Myopic\_Bias $\circ$$\rightarrow$  Income ~& 0.1\\ \hline
Myopic\_Bias $\circ$–$\circ$ Income ~& 0.06\\ \hline
Myopic\_Bias $\rightarrow$ Income ~& 0.05\\ \hline
Myopic\_Bias $\rightarrow$ Planning ~& 0.03\\ \hline
Myopic\_Bias $\leftrightarrow$ Planning ~& 0.01\\ \hline
Planning $\circ$$\rightarrow$  Asset\_Amt ~& 0.1\\ \hline
Planning $\rightarrow$ Asset\_Amt ~& 0.08\\ \hline
Planning $\rightarrow$ Confidence ~& 0.01\\ \hline
Planning $\rightarrow$ Fin\_Literacy ~& 0.27\\ \hline
Planning $\circ$$\rightarrow$  Fin\_Literacy ~& 0.19\\ \hline
Planning $\circ$$\rightarrow$  Income ~& 0.06\\ \hline
Planning $\rightarrow$ Income ~& 0.01\\ \hline
Planning $\rightarrow$ Myopic\_Bias ~& 0.01\\ \hline
Planning $\circ$$\rightarrow$  Myopic\_Bias ~& 0.01\\ \hline

\end{supertabular}

\vspace{20pt} 

\tablefirsthead{\hline
    \multicolumn{2}{|c|}{\textbf{Group 8}} \\
    \hline
    \textbf{Edge} & \textbf{Prob.} \\ 
    \hline
}
\tablehead{\hline
    \multicolumn{2}{|c|}{\textbf{Group 8 (Continued)}} \\
    \hline
    \textbf{Edge} & \textbf{Prob.} \\
    \hline
}
\tabletail{
    \hline
    \multicolumn{2}{|r|}{\textit{To be Continued}} \\
    \hline
}
\tablelasttail{\hline}
\begin{supertabular}{|p{5cm}|p{1cm}|}
Age $\rightarrow$ Asset\_Amt ~& 1\\ \hline
Age $\rightarrow$ Fin\_Literacy ~& 1\\ \hline
Age $\rightarrow$ Herding\_Bias ~& 0.89\\ \hline
Age $\rightarrow$ Invest ~& 0.88\\ \hline
Age $\rightarrow$ Myopic\_Bias ~& 1\\ \hline
Asset\_Amt $\rightarrow$ Confidence ~& 0.47\\ \hline
Asset\_Amt $\rightarrow$ Fin\_Literacy ~& 0.77\\ \hline
Asset\_Amt $\rightarrow$ Invest ~& 0.7\\ \hline
Asset\_Amt $\leftrightarrow$ Invest ~& 0.05\\ \hline
Asset\_Amt $\circ$$\rightarrow$  Myopic\_Bias ~& 0.28\\ \hline
Asset\_Amt $\rightarrow$ Myopic\_Bias ~& 0.25\\ \hline
Asset\_Amt $\rightarrow$ Planning ~& 0.49\\ \hline
Asset\_Amt $\leftrightarrow$ Planning ~& 0.24\\ \hline
Confidence $\rightarrow$ Asset\_Amt ~& 0.4\\ \hline
Confidence $\leftrightarrow$ Asset\_Amt ~& 0.13\\ \hline
Confidence $\rightarrow$ Fin\_Literacy ~& 0.32\\ \hline
Confidence $\circ$$\rightarrow$  Fin\_Literacy ~& 0.31\\ \hline
Confidence $\leftrightarrow$ Fin\_Literacy ~& 0.29\\ \hline
Confidence $\leftrightarrow$ Income ~& 0.22\\ \hline
Confidence $\rightarrow$ Income ~& 0.02\\ \hline
Confidence $\circ$$\rightarrow$  Income ~& 0.01\\ \hline
Confidence $\rightarrow$ Invest ~& 0.4\\ \hline
Confidence $\leftrightarrow$ Invest ~& 0.23\\ \hline
Confidence $\circ$$\rightarrow$  Invest ~& 0.05\\ \hline
Confidence $\circ$–$\circ$ Invest ~& 0.01\\ \hline
Confidence $\leftrightarrow$ Planning ~& 0.39\\ \hline
Confidence $\rightarrow$ Planning ~& 0.3\\ \hline
Education $\rightarrow$ Confidence ~& 0.99\\ \hline
Education $\rightarrow$ Fin\_Literacy ~& 1\\ \hline
Education $\rightarrow$ Income ~& 1\\ \hline
Education $\rightarrow$ Invest ~& 0.81\\ \hline
Education $\rightarrow$ Myopic\_Bias ~& 1\\ \hline
Fin\_Literacy $\rightarrow$ Asset\_Amt ~& 0.23\\ \hline
Fin\_Literacy $\rightarrow$ Confidence ~& 0.08\\ \hline
Fin\_Literacy $\leftrightarrow$ Invest ~& 0.31\\ \hline
Fin\_Literacy $\rightarrow$ Invest ~& 0.01\\ \hline
Fin\_Literacy $\rightarrow$ Myopic\_Bias ~& 0.04\\ \hline
Fin\_Literacy $\leftrightarrow$ Planning ~& 0.13\\ \hline
Fin\_Literacy $\rightarrow$ Planning ~& 0.01\\ \hline
Herding\_Bias $\leftrightarrow$ Confidence ~& 0.2\\ \hline
Herding\_Bias $\rightarrow$ Confidence ~& 0.01\\ \hline
Herding\_Bias $\circ$$\rightarrow$  Confidence ~& 0.01\\ \hline
Herding\_Bias $\leftrightarrow$ Fin\_Literacy ~& 0.54\\ \hline
Herding\_Bias $\circ$$\rightarrow$  Fin\_Literacy ~& 0.46\\ \hline
Herding\_Bias $\leftrightarrow$ Income ~& 0.47\\ \hline
Herding\_Bias $\rightarrow$ Invest ~& 0.01\\ \hline
Income $\rightarrow$ Asset\_Amt ~& 0.52\\ \hline
Income $\leftrightarrow$ Asset\_Amt ~& 0.48\\ \hline
Income $\circ$$\rightarrow$  Confidence ~& 0.11\\ \hline
Income $\rightarrow$ Confidence ~& 0.01\\ \hline
Income $\circ$$\rightarrow$  Fin\_Literacy ~& 0.03\\ \hline
Income $\leftrightarrow$ Invest ~& 0.01\\ \hline
Income $\circ$$\rightarrow$  Myopic\_Bias ~& 0.28\\ \hline
Income $\rightarrow$ Myopic\_Bias ~& 0.07\\ \hline
Income $\rightarrow$ Planning ~& 0.52\\ \hline
Income $\leftrightarrow$ Planning ~& 0.34\\ \hline
Invest $\rightarrow$ Asset\_Amt ~& 0.25\\ \hline
Invest $\rightarrow$ Confidence ~& 0.3\\ \hline
Invest $\circ$$\rightarrow$  Confidence ~& 0.01\\ \hline
Invest $\circ$$\rightarrow$  Fin\_Literacy ~& 0.62\\ \hline
Invest $\rightarrow$ Fin\_Literacy ~& 0.06\\ \hline
Myopic\_Bias $\rightarrow$ Asset\_Amt ~& 0.25\\ \hline
Myopic\_Bias $\leftrightarrow$ Asset\_Amt ~& 0.21\\ \hline
Myopic\_Bias $\leftrightarrow$ Confidence ~& 0.01\\ \hline
Myopic\_Bias $\rightarrow$ Fin\_Literacy ~& 0.47\\ \hline
Myopic\_Bias $\leftrightarrow$ Fin\_Literacy ~& 0.46\\ \hline
Myopic\_Bias $\circ$$\rightarrow$  Fin\_Literacy ~& 0.03\\ \hline
Myopic\_Bias $\leftrightarrow$ Income ~& 0.52\\ \hline
Myopic\_Bias $\rightarrow$ Income ~& 0.01\\ \hline
Myopic\_Bias $\leftrightarrow$ Planning ~& 0.63\\ \hline
Planning $\rightarrow$ Asset\_Amt ~& 0.21\\ \hline
Planning $\circ$$\rightarrow$  Asset\_Amt ~& 0.06\\ \hline
Planning $\rightarrow$ Confidence ~& 0.26\\ \hline
Planning $\circ$$\rightarrow$  Confidence ~& 0.05\\ \hline
Planning $\rightarrow$ Fin\_Literacy ~& 0.81\\ \hline
Planning $\circ$$\rightarrow$  Fin\_Literacy ~& 0.05\\ \hline
Planning $\rightarrow$ Income ~& 0.12\\ \hline
Planning $\circ$$\rightarrow$  Income ~& 0.02\\ \hline
Planning $\leftrightarrow$ Invest ~& 0.16\\ \hline
Planning $\rightarrow$ Myopic\_Bias ~& 0.23\\ \hline
Planning $\circ$$\rightarrow$  Myopic\_Bias ~& 0.09\\ \hline

\end{supertabular}

\end{document}